\documentclass[final,3p,times,twocolumn]{elsarticle}

\usepackage{amssymb}
\usepackage{amsmath}

\usepackage{lineno}

\usepackage{pifont}
\newcommand{\cmark}{\ding{51}}%
\newcommand{\xmark}{\ding{55}}%
\usepackage{blindtext}
\usepackage{cuted}
\usepackage{caption}
\usepackage{hyperref}
\RequirePackage[capitalise]{cleveref}
\usepackage{mymacros}
\usepackage[ruled,vlined]{algorithm2e}
\RestyleAlgo{ruled}
\SetKwComment{Comment}{/* }{*/}

\usepackage{mathtools}
\usepackage{nicefrac}
\usepackage{color,soul}
\newcounter{problem}
\newtheorem{Problem}[problem]{Problem}
\usepackage{multirow}
\usepackage{xcolor, colortbl}

\newcommand{\arraystretchfortable}{1.3}

\newcommand{\tablerowsep}{3.5ex}

\usepackage{tablefootnote}

\journal{Acta Astronautica}

\begin{document}
\begin{frontmatter}



\title{Delta-V-Optimal Centralized Guidance Strategy For Under-actuated N-Satellite Formations}


\author[1]{Ahmed Mahfouz}
\author[2]{Gabriella Gaias}
\author[3]{Florio {Dalla Vedova}}
\author[1,4]{Holger Voos}

\affiliation[1]{organization={SnT, University of Luxembourg},
            addressline={29, Avenue J.F Kennedy}, 
            city={Luxembourg},
            postcode={1855}, 
            country={Luxembourg}}

\affiliation[2]{organization={Department of Aerospace Science and Technology, Politecnico di Milano},
            addressline={34, via La Masa}, 
            city={Milan},
            postcode={20156}, 
            country={Italy}}

\affiliation[3]{organization={LuxSpace},
            addressline={9, Rue Pierre Werner}, 
            city={Betzdorf},
            postcode={6832}, 
            country={Luxembourg}}

\affiliation[4]{organization={Faculty of Science, Technology and Medicine, University of Luxembourg},
            addressline={2, place de l’Université}, 
            city={Esch-sur-Alzette},
            postcode={4365}, 
            country={Luxembourg}}

\begin{abstract}
This paper addresses the computation of Delta-V-optimal, safe, relative orbit reconfigurations for satellite formations in a centralized fashion. The formations under consideration comprise an uncontrolled chief spacecraft flying with  
an arbitrary number, $N$, of deputy satellites, where each deputy is equipped with a single electric thruster.
Indeed, this represents a technological solution that is becoming widely employed by the producers of small-satellite platforms. While adopting a single electric thruster does reduce the required power, weight, and size of the orbit control system, it comes at the cost of rendering the satellite under-actuated. In this setting, the satellite can provide a desired thrust vector only after an attitude maneuver is carried out to redirect the thruster nozzle opposite to the desired thrust direction.
In order to further extend the applicability range of such under-actuated platforms, guidance strategies are developed to support different reconfiguration scenarios for $N$-satellite formations.
This paper starts from a classical non-convex quadratically constrained trajectory optimization formulation, which passes through multiple simplifications and approximations to arrive to two novel convex formulations, namely a second-order cone programming formulation, and a linear programming one. Out of five guidance formulations proposed in this article, the most promising three were compared through an extensive benchmark analysis that is applied to fifteen of the most widely-used solvers. This benchmark experiment provides information about the key distinctions between the different problem formulations, and under which conditions each one of them can be recommended.

\end{abstract}




\end{frontmatter}



\section{Introduction}
As the world entered the new space era, it has witnessed a shift in the design philosophy of satellites. A modern space mission commonly comprises one or more simpler, smaller, lighter, lower-powered, and cheaper satellites, in contrast to a traditional mission, which typically relied on a single, sophisticated, large, heavy, power demanding, and expensive spacecraft \cite{Pelton2020SmallSatellites}. By deploying multiple small satellites in a coordinated formation, these missions can cover larger areas, provide redundancy, and offer more frequent data updates.\\

As the satellites gradually became smaller in size, it was no wonder that Electric Propulsion (EP) systems were increasingly used to support their maneuverability in orbit. At the cost of lower thrust levels, electric thrusters typically offer higher specific impulse, and hence better fuel efficiency than their chemical counterparts \cite{Miller2021Survey}. The low fuel and propellant storage requirements is mainly what makes EP an attractive option for small satellites \cite{OReilly2021EP}.
Often motivated by the need to reduce weight, complexity, and hence cost, some satellite manufacturers started opting to incorporate a single electric thruster onboard their spacecraft. Examples of such satellites include the PLATiNO platform \cite{Stanzione2018PLATiNO} and Triton-X \cite{Helmeid2022TheIntegrated}. A satellite that employs a single electric thruster is, in nature, under-actuated, since a desired thrust vector is only achievable after an attitude maneuver which redirects the thruster nozzle into the desired direction. This paper focuses on the problem of optimizing the trajectories necessary for a reconfiguration of an arbitrary number of such under-actuated satellites flying together in a formation.\\

The problem of formation reconfiguration has been a subject of extensive research. While guidance and control strategies have been proposed for formations that leverage impulsive thrusters \cite{Larsson2011Autonomous, Gaias2015Impulsive, di2018continuous, chernick2018new}, they are not applicable to formations that employ electric ones. Existing low-thrust guidance schemes assume omnidirectional thrusting \cite{scala2021design, DeVittori2022Low-Thrust}, rendering them inapplicable to single-thruster satellite formations. Guidance and control policies were developed for the AVANTI mission \cite{Gaias2018Avanti, Gaias2015Impulsive, Gaias2015Generalized}, which incorporates a satellite equipped with one impulsive thruster flying in a formation with a non-cooperative object, yet again, these guidance and control methods are not suitable for low-thrust formations. On a related note, a satellite incorporating a single impulsive thruster cannot be considered under-actuated, unlike a spacecraft with a sole low-thruster. An impulsive thruster is idle for the most part of the maneuver due to the fact that it is able to provide large Delta-V changes, typically in a matter of seconds. A thruster-idle period is ample for an attitude redirection maneuver to take place before the following thruster firing. To address the problem of relative orbit corrections for mono-electric-thruster satellites, Model Predictive Control (MPC) schemes have been proposed in \cite{Belloni2023Relative, Mahfouz2023Autonomous}, yet operational constraints, such as the necessary thruster-off-periods during ground contact, during payload operations, or during eclipse, were not considered in these works.\\

In this article, the Delta-V-optimal, collision-free, formation reconfiguration problem is considered for an arbitrary number, $N$, of deputy satellites flying in a formation with an uncontrolled chief. To allow for such reconfiguration to take place, each deputy is equipped with one electric thruster. The proposed trajectory optimization procedures are open-loop control strategies which can be incorporated in the closed feedback loop in many ways; one of which is a shrinking-horizon MPC such as the one presented in \cite{Mahfouz2024Fuel-Optimal}.
The first attempt to formulate the guidance problem is presented as a non-convex Quadratically Constrained Quadratic Programming (QCQP) problem, which is later relaxed to a convex QCQP that can be solved using Sequential Convex Programming (SCP). Due to the fact that the optimal solutions to the QCQP formulations require unnecessarily large Delta-V changes, the convex QCQP problem is reformulated into a Second-Order Cone Programming (SOCP) problem which makes better use of the available fuel. A final relaxation is applied to the SOCP formulation to transform it into the simplest form of convex programming problems; Linear Programming (LP). 
The main contributions of this paper are as follows:
\begin{itemize}
    \item The mathematical formulations of the SOCP and the LP problems, which produce more Delta-V efficient solutions in comparison to the classical QCQP problems, while at the same time being easier to implement and faster to solve by most solvers;
    \item A benchmark experiment which is carried out on more than a dozen solvers over the convex QCQP, the SOCP, and the LP formulations;
    \item Guidelines for embedding the guidance algorithms for space-borne applications, drawn considering typical reconfigurations problems tackled in remote sensing applications.
\end{itemize}

The guidance plans presented in this paper are implemented as part of the AuFoSat toolbox; a Guidance, Navigation and Control library, developed to support the future missions of Triton-X; the multi-mission Low-Earth Orbit (LEO) microsatellite platform developed by LuxSpace. Previous AuFoSat research discussed orbit design \cite{menzio2022formation}, relative navigation \cite{mahfouz2022relative, Mahfouz2023GNSS-based}, absolute orbit keeping \cite{Mahfouz2023Autonomous}, and relative orbit corrections for two-satellite formations \cite{Mahfouz2024Fuel-Optimal}.\\

The article is organized such that \cref{sec:Dynamics} introduces the mathematical model used in the development of the trajectory optimization formulations, namely, a linearized model of the quasi-non-singular Relative Orbital Elements (ROE) is introduced. In \cref{sec:Guidance}, the different formulations of the formation reconfiguration problem are presented. \cref{sec:Results_and_discussion} contains the implications of employing the final LP formulation, and presents the benchmark experiment. It also discusses the limitations of the proposed guidance schemes. Lastly, the paper is concluded in \cref{sec:Conclusion}.

\section{Dynamical model}\label{sec:Dynamics}
In this section, a dynamical model of the quasi-non-singular Relative Orbital Elements (ROE), which is necessary for the development of the trajectory optimization strategies, is introduced. Multiple reference frames are utilised in the development of the ROE dynamics. The ones used in this work are; an Earth-Centered-Inertial frame (ECI), denoted as $\set{F}^{i}$. The True of Date (TOD) Earth Equator frame is the one referred to by ECI in this article; a Satellite-body-fixed frame, denoted by $\set{F}^{b}$, with axes along the principal axes of inertia of the satellite; and the Radial-Transversal-Normal frame (RTN), denoted as $\set{F}^{r}$, with its x-axis along the position vector of the chief pointing away from the Earth, with its z-axis along the normal direction to the chief's orbital plane, and with its y-axis completing the right-handed triad.
Vectors expressed in $\set{F}^{i}$, $\set{F}^{b}$, or $\set{F}^{r}$ are signified by the superscripts $\parenth{\cdot}^{i}$, $\parenth{\cdot}^{b}$, or $\parenth{\cdot}^{r}$ respectively.\\

The following set of orbital elements can be used to describe the orbital motion of a satellite, which is under the gravitational influence of a large celestial body such as the Earth, in a planet centered reference frame,
\begin{equation} \label{eq:OE}
    \vec{\alpha} \coloneqq \begin{bmatrix} a & \theta & e_{x} & e_{y} & i & \Omega \end{bmatrix}^{\intercal}, 
\end{equation}
where $a$ is the semi-major axis, $\theta$ is the mean argument of latitude, $\vec{e}\coloneqq\begin{bmatrix} e_{x} & e_{y} \end{bmatrix}^{\intercal} = \begin{bmatrix} e \cos{\omega} & e\sin{\omega} \end{bmatrix}^{\intercal}$, is the eccentricity vector with $e$ being the eccentricity of the orbit and $\omega$ being the argument of periapsis. Furthermore, $i$ is the orbital inclination and $\Omega$ is the Right Ascension of the Ascending Node (RAAN). A Cartesian state vector, $\vec{x}^{i}\coloneqq \begin{bmatrix} \parenth{\vec{r}^{i}}^{\intercal} & \parenth{\vec{v}^{i}}^{\intercal}
\end{bmatrix}^{\intercal}$, can equally represent the motion of the satellite, where $\vec{r}^{i}$ and $\vec{v}^{i}$ are the absolute position and velocity vectors expressed in $\set{F}^{i}$. Note that the Cartesian state vector can be mapped through a nonlinear transformation to the classical orbital elements, and vice versa \cite{vallado2001fundamentals}. When this mapping is applied to the precise Cartesian state vector of the satellite, the instantaneous osculating orbital elements are produced, which, in the discussions to follow, are denoted by $\tilde{\vec{\alpha}}$. Conversely, the mean orbital elements, denoted in the rest of the paper by $\vec{\alpha}$, are the one-orbit averaged elements, where the short- and long-term oscillations generated by the $J_{2}$ harmonic of the Earth gravitational potential are removed. The transformations in \cite{Gaias2020Analytical} are utilized to perform the mean/osculating elements mapping.\\
  
The motion of the $i^\text{th}$ deputy satellite with respect to the chief spacecraft is parameterized in this research by the dimensionless quasi-non-singular ROE vector, which is defined, using the orbital elements of both, the deputy and the chief, as follows,
\begin{equation} \label{eq:ROE}
    \delta_{i} \vec{\alpha} \coloneqq \begin{bmatrix} \delta_{i} a \\ \delta_{i} \lambda \\ \delta_{i} e_{x} \\ \delta_{i} e_{y} \\ \delta_{i} i_{x} \\ \delta_{i} i_{y} \end{bmatrix} = \begin{bmatrix} \Delta_{i} a/a_{c} \\ \Delta_{i} \theta + \Delta_{i} \Omega \cos{i_{c}} \\ \Delta_{i} e_{x} \\ \Delta_{i} e_{y} \\ \Delta_{i} i \\ \Delta_{i} \Omega \sin{i_{c}} \end{bmatrix},
\end{equation}
where $\delta_{i} \vec{\alpha}$ is the dimensionless ROE vector of the $i^\text{th}$ deputy, $\delta_{i} a$ is its relative semi-major axis, $\delta_{i} \lambda$ is the relative mean longitude, $\delta_{i}\vec{e}\coloneqq\begin{bmatrix} \delta_{i} e_{x} & \delta_{i} e_{y}\end{bmatrix}^{\intercal}$ is the relative eccentricity vector, and $\delta_{i}\vec{i}\coloneqq\begin{bmatrix} \delta_{i} i_{x} & \delta_{i} i_{y}\end{bmatrix}^{\intercal}$ is the relative inclination vector. It is to be noted that in this paper, the subscript $\parenth{\cdot}_{i} \;\forall i \in \left\{1, 2, \hdots, N\right\}$, with $N$ being the number of deputies, denotes a quantity related to the $i^\text{th}$ deputy satellite, while the subscript $\parenth{\cdot}_{c}$ is used for chief-related quantities. Furthermore, $\delta_{i} \parenth{\cdot}$ signifies a relative quantity between the $i^\text{th}$ deputy and the chief, which may or may not be the arithmetic difference between the two quantities, while $\Delta_{i} \parenth{\cdot}$ signifies the arithmetic difference between $\parenth{\cdot}_{i}$ and $\parenth{\cdot}_{c}$, i.e.,  $\Delta_{i} \parenth{\cdot} \coloneqq \parenth{\cdot}_{i} - \parenth{\cdot}_{c}$. As in the case of absolute orbital elements, the osculating ROE vector is denoted by $\delta_{i} \tilde{\vec{\alpha}}$, whereas the mean ROE vector is referred to as $\delta_{i} {\vec{\alpha}}$. A dimensional ROE vector is obtained by multiplying the dimensionless ROE vector by the semi-major axis of the chief,
\begin{equation} \label{eq:y_def}
    \vec{y}_{i} = a_{c} \delta_{i} \vec{\alpha},
\end{equation}
where $\vec{y}_{i}$ is the dimensional mean ROE vector of the $i^\text{th}$ deputy, with units of length.\\

Assuming neighbouring orbits of the chief and the deputies, and a near-circular orbit of the chief, the dynamics of the ROE can be linearized to the first order considering the mean effect of the $J_{2}$ zonal harmonic. In fact, a closed form solution of the linearized dynamics can be obtained for piece-wise constant input acceleration as discussed in \cite{di2018continuous}. The system evolution is expressed in the following form,
\begin{equation}\label{eq:ROE_dynamics_sol}
    \vec{y}_{i} \parenth{t_{k+1}} = \mat{\Phi}\parenth{t_{k}, t_{k+1}} \vec{y}_{i} \parenth{t_{k}} + \mat{\Psi}\parenth{t_{k}, t_{k+1}} \bar{\vec{u}}_{i}^{r}\parenth{t_{k}, t_{k+1}}, 
\end{equation}
where $\mat{\Phi}\parenth{t_{k}, t_{k+1}}$ is the State Transition Matrix (STM) between the two time instants, $t_{k}$ and $t_{k+1}$, $\mat{\Psi}\parenth{t_{k}, t_{k+1}}$ is the convolution matrix between the same two time instants, and $\bar{\vec{u}}_{i}^{r}\parenth{t_{k}, t_{k+1}} = a_{c} \vec{u}_{i}^{r}\parenth{t_{k}, t_{k+1}}$, with $\vec{u}_{i}^{r}\parenth{t_{k}, t_{k+1}} = \begin{bmatrix} u_{i, R}\parenth{t_{k}, t_{k+1}} & u_{i, T}\parenth{t_{k}, t_{k+1}} & u_{i, N}\parenth{t_{k}, t_{k+1}} \end{bmatrix}^{\intercal}$ being the input acceleration vector provided by the $i^\text{th}$ deputy's thruster, expressed in $\set{F}^{r}$, and constant over the period $\left[t_{k}, t_{k+1}\right)$. In the rest of the text, and in order to simplify the representation of equations, the following notations are used, $\mat{\Phi}_{k} \equiv \mat{\Phi}\parenth{t_{k}, t_{k+1}}$, $ \mat{\Psi}_{k} \equiv \mat{\Psi}\parenth{t_{k}, t_{k+1}}$, $\vec{y}_{k} \equiv \vec{y}\parenth{t_{k}}$, $\bar{\vec{u}}_{i, k} \equiv \bar{\vec{u}}_{i}^{r}\parenth{t_{k}, t_{k+1}}$, and $\vec{u}_{i, k} \equiv \vec{u}_{i}^{r}\parenth{t_{k}, t_{k+1}}$. 
Note that $\mat{\Phi}_{k}$ and $\mat{\Psi}_{k}$ do not relate to the $i^\text{th}$ deputy since they only depend on chief-related quantities \cite{di2018continuous}.

\section{Guidance}\label{sec:Guidance}
\begin{figure*}[ht]
    \centering
    \includegraphics[width=\linewidth]{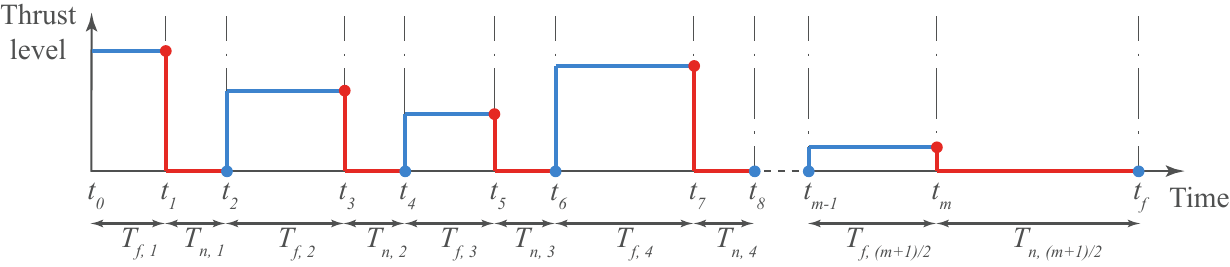}
    \caption{Graphical representation of the low-thrust guidance strategy}
    \label{fig:Low-thrust-guidance-scheme}
\end{figure*}

In this section, a guidance scheme is developed for a formation that comprises an arbitrary number, $N$, of deputy satellites together with a chief spacecraft. The ultimate goal of the guidance module is to drive the state of each deputy from its initial values, $\vec{y}_{i, 0}$ at the initial time, $t_{0}$, to the user-defined reference state, $\vec{y}_{i, f}$ at the final time, $t_{f}$.
The $N$ deputies are assumed to be each equipped with a single throttleable electric thruster dedicated for (relative) orbit reconfiguration, while the chief is uncontrolled and is assumed to be the spacecraft onboard of which the guidance calculations take place. 
In this under-actuated setting, each deputy has to undergo repeating attitude maneuvers in order to redirect its thruster in the desired direction before every thruster firing. The time required for these redirection maneuvers was taken into consideration in the design phase of the guidance procedure, and the thruster of each deputy is thus dictated to operate on an alternating on/off mode, where the slew maneuvers are allocated in the thruster off-periods. On top of the natural relative orbit changes during the thrusters off-periods, further forced ROE corrections are simultaneously induced for all the deputies through $\dfrac{m+1}{2}$ finite burns, with $m$ being an odd number.
Figure \ref{fig:Low-thrust-guidance-scheme} illustrates the alternation between the on and off states of the electric thruster onboard of one of the deputies which shall drive its state from the initial to the reference value.

It is important to emphasize that the time instances at which the thruster alternates between the on and off states, $\vec{t} = \begin{bmatrix} t_{0} & t_{1} & \hdots & t_{f}
\end{bmatrix}$ (see \cref{fig:Low-thrust-guidance-scheme}), are assumed to be identical for all the deputies. It is, in fact, a common practice to unify the discretization steps for all the deputies in centralized trajectory optimization schemes \cite{scala2021design, Morgan14MPC}. 
The problem of specifying a separate time vector for each deputy is out of the scope of this paper, and will be the focus of our future research. In fact, the problems of computing and broadcasting this time vector are also out of the scope of this study. The common time vector is left for the mission operator to determine, in order to enhance the mission predictability, which is a very important factor in the operation of real missions. It appears that leaving $\vec{t}$ as a user-input not only helps predict the behaviour of the formation ahead of time, but also helps accommodate any operational constraints within the thrusters idle intervals. Operational constraints in our context might include not being able to use the thruster or the attitude control system for relative orbit correction during specific times, e.g., during ground contact, payload operations, or during times when the satellite is in the shadow. The time vector, $\vec{t}$, can be defined either through the time instances at which the thrusters switch between the on and off states, or through defining the initial time, $t_{0}$, and the periods during which the thrusters are turned on or off. As a result, by letting $\dist{L} = \curlyb{1, 2, \hdots, \parenth{m+1}/2}$, these periods are divided into two categories; 
\begin{itemize}
    \item The forced motion time periods, $T_{f, l}$, $l \in \dist{L}$; 
    \item The natural motion periods (coast arcs), $T_{n, l}$, $l \in \dist{L}$.
\end{itemize} 
Since the attitude redirection maneuvers are allocated during the natural motion periods, a lower bound for these coast arcs has to be imposed which is related to the maximum allowable angular speed, i.e., angular velocity's Euclidean norm, of the deputies, such that,
\begin{equation}\label{eq:Tn}
    T_{n,l} \geq \frac{\pi}{\omega_{max}} + T_\text{safety} \quad \forall l \in \dist{L},
\end{equation}
where $\omega_{max}=\min\parenth{\omega_{i, max}}$ is the maximum angular speed of the  deputy with slowest angular rate, with $\omega_{i, max}$ being the maximum angular speed of the $i^\text{th}$ deputy, and $T_\text{safety} \geq 0$ is added to ensure that the coast arc can accommodate the most stringent attitude maneuver.\\

Letting,
\begin{equation}\label{eq:Y_def}
    \mat{Y} = \begin{bmatrix}
        \vec{y}_{1, 0} & \vec{y}_{2, 0} & \hdots & \vec{y}_{N, 0} \\
        \vec{y}_{1, 1} & \vec{y}_{2, 1} & \hdots & \vec{y}_{N, 1} \\
        \vdots & \vdots & \ddots & \vdots \\
        \vec{y}_{1, m+1} & \vec{y}_{2, m+1} & \hdots & \vec{y}_{N, m+1}
    \end{bmatrix},
\end{equation}
\begin{equation}\label{eq:DV_def}
    \bar{\mat{U}} = \begin{bmatrix}
        \bar{\vec{u}}_{1, 0} & \bar{\vec{u}}_{2, 0} & \hdots & \bar{\vec{u}}_{N, 0} \\
        \bar{\vec{u}}_{1, 1} & \bar{\vec{u}}_{2, 1} & \hdots & \bar{\vec{u}}_{N, 1} \\
        \vdots & \vdots & \ddots & \vdots \\
        \bar{\vec{u}}_{1, m} & \bar{\vec{u}}_{2, m} & \hdots & \bar{\vec{u}}_{N, m} \end{bmatrix},
\end{equation}
the guidance problem can be formally written as an optimization problem with a quadratic objective function, as is classically done \cite{pippia2022reconfiguration}, as follows,
\begin{Problem}[Non-convex formulation]
\label{prob:nonconvex_formulation}
\begin{align}
& \min_{\mat{Y}, \bar{\mat{U}}} \quad \frac{1}{a_{c}^{2}}\sum_{i\in \dist{I}} \sum_{k \in \dist{K}_{f}}{\parenth{\Delta t_{k}^{2} \bar{\vec{u}}_{i, k}^{\intercal} \bar{\vec{u}}_{i, k}}} \nonumber\\
& \text{subject to,} \nonumber\\
& \vec{y}_{i, 0} = \vec{y}_{i, 0}, \qquad \vec{y}_{i, m+1} = \vec{y}_{i, f} \quad \forall i \in \dist{I}, \label{eq:boundary_constraints_nonconvex}\\
& \vec{y}_{i, k+1} =  \mat{\Phi}_{k} \vec{y}_{i, k} + \mat{\Psi}_{k} \bar{\vec{u}}_{i, k} \quad \forall i \in \dist{I} ,\; \forall k \in \dist{K},\label{eq:dynamics_constraint_nonconvex}\\
& \bar{\vec{u}}_{i, k} = \vec{0} \quad \forall i \in \dist{I} ,\; \forall k \in \dist{K}_{n}, \label{eq:u0_constraint_nonconvex}\\
& \bar{\vec{u}}^{\intercal}_{i, k}\bar{\vec{u}}_{i, k} \leq a_{c}^{2} u^{2}_{i, max} \quad \forall i \in \dist{I} ,\; \forall k \in \dist{K}_{f}, \label{eq:umax_constraint_nonconvex}\\
& \begin{multlined}
    \parenth{\vec{y}_{i, k} - \vec{y}_{j, k}}^{\intercal} \mat{T}_{k}^{\intercal} \mat{T}_{k} \parenth{\vec{y}_{i, k} - \vec{y}_{j, k}} \geq R_\text{CA}^{2}\\
    \hfill \forall i, j \in \dist{I} ,\; i \neq j ,\; \forall k \in \dist{K}, \label{eq:CA_deputy_deputy_nonconvex}
\end{multlined}\\
& \bar{\vec{y}}_{i, k}^{\intercal} \mat{T}_{k}^{\intercal} \mat{T}_{k} \vec{y}_{i, k} \geq R_\text{CA}^{2} \quad \forall i \in \dist{I} ,\; \forall k \in \dist{K},\label{eq:CA_deputy_chief_nonconvex}
\end{align}
\end{Problem}
where $\Delta t_{k} = t_{k+1} - t_{k}$, $\vec{0}$ is a vector of zeros, $\dist{I} = \curlyb{1, 2, \hdots, N}$ is the set of deputies' indices, $\dist{K} = \dist{K}_{f} \cup \dist{K}_{n}$, with $\dist{K}_{f} = \curlyb{0, 2, 4, \hdots, m-1}$ being the set of time indices where the thruster of each deputy is turned on, i.e., forced motion periods, and $\dist{K}_{n} = \left\{1, 3, 5 \hdots, m\right\}$ being the set of time indices where the thruster of each deputy is turned off, i.e., natural motion periods, $u_{i, max}$ is the maximum allowable acceleration by the onboard thruster of the $i^\text{th}$ deputy. Generally, it is the maximum allowable thrust that is provided in the data sheets of electric thrusters, and not the maximum acceleration. However, the latter can be calculated from the former according to: $u_{i, max} = f_{i, max}/M_{i}$, with $f_{i, max}$ being the maximum thrust of the $i^\text{th}$ deputy, and $M_{i}$ being its mass, which is assumed constant throughout the maneuver, and is set to the mass of the satellite at $t_{0}$. Furthermore, in Problem \ref{prob:nonconvex_formulation}, $R_\text{CA} \geq 0$ is the radius of the collision avoidance sphere, and $\mat{T}_{k} \in \set{R}^{3\times 6}$ is the matrix that transforms a dimensional ROE vector into its corresponding position vector in the RTN frame. An explicit expression for this matrix can be found in \cite{Gaias2021Trajectory}. It is important to note that $\bar{\vec{u}}_{i,k}$ are chosen to be included as optimization variables instead of $\vec{u}_{i,k}$ in order to ensure that all the decision variables, i.e., $\mat{Y}$ and $\bar{\mat{U}}$, are of comparable orders of magnitude, which makes it less probable that a solver will run into numerical issues. Poorly-scaled problems typically require longer times to solve, if a solution can be found in the first place. \\

The constraints imposed on Problem \ref{prob:nonconvex_formulation} are summarized as follows,
\begin{itemize}
    \item Equation \eqref{eq:boundary_constraints_nonconvex} represents the boundary constraints which dictates the guidance profile to have the final state exactly equal to the set points defined by the user, while respecting the initial state of each deputy;
    \item The relative orbital dynamics are imposed on the trajectory optimizer through \cref{eq:dynamics_constraint_nonconvex};
    \item The acceleration constraints are forced through equations \eqref{eq:u0_constraint_nonconvex} and \eqref{eq:umax_constraint_nonconvex}.  Note that \cref{eq:u0_constraint_nonconvex} is a hard constraint to ensure that the input acceleration onboard of each deputy provided during attitude redirection maneuvers is exactly zero;
    \item Inter-deputy collision is avoided by imposing \cref{eq:CA_deputy_deputy_nonconvex}, which guarantees that no deputy enters the collision sphere of another, while deputy-chief collision is prohibited at each time step through \cref{eq:CA_deputy_chief_nonconvex}. Since the time steps of our application are relatively small when considering how slow the relative orbital dynamics can be, inter-step collision avoidance is ignored in this study.
\end{itemize}

It is clear that the cost function of Problem \ref{prob:nonconvex_formulation} is the sum of the squared second Lebesgue (L2) norms of all the control Delta-V across all deputies and across all time instances during the reconfiguration maneuver. Minimizing this cost function not only results in a Delta-V-optimal solution, but also produces a fuel-optimal solution. Indeed, the terms "Delta-V-optimal" and "fuel-optimal" can be used interchangeably in our case, since they really do refer to the same thing when the controlled spacecraft are each equipped with a single thruster \cite{ROSS2006SpaceTrajectoryOptimization}.\\

Problem \ref{prob:nonconvex_formulation} is a non-convex optimization problem which requires a series of elaborate processes in order to find its globally optimal solution. The non-convexity of Problem \ref{prob:nonconvex_formulation} arises solely from the non-convexity of the two last constraints. In fact, if it were not for constraints \eqref{eq:CA_deputy_deputy_nonconvex} and \eqref{eq:CA_deputy_chief_nonconvex}, Problem \ref{prob:nonconvex_formulation} would have been a convex optimization problem that is guaranteed to have a globally optimal solution. One way to convexify the problem is to approximate the non-convex constraints by affine ones, then solving the problem through Sequential Convex Programming (SCP). Using the relaxation proposed in \cite{Morgan14MPC} for the collision avoidance constraints, Problem \ref{prob:nonconvex_formulation} can be rewritten in its relaxed convex Quadratically Constrained Quadratic Programming (convex QCQP) form as follows,
\begin{Problem}[Convex QCQP formulation]
\label{prob:QCQP_formulation}
\begin{align}
& \min_{\mat{Y}, \bar{\mat{U}}} \quad \frac{1}{a_{c}^{2}}\sum_{i\in \dist{I}} \sum_{k \in \dist{K}_{f}}{\parenth{\Delta t_{k}^{2} \bar{\vec{u}}_{i, k}^{\intercal} \bar{\vec{u}}_{i, k}}} \nonumber\\
& \text{subject to,} \nonumber\\
& \vec{y}_{i, 0} = \vec{y}_{i, 0}, \qquad \vec{y}_{i, m+1} = \vec{y}_{i, f} \quad \forall i \in \dist{I}, \label{eq:boundary_constraints_QCQP}\\
& \vec{y}_{i, k+1} =  \mat{\Phi}_{k} \vec{y}_{i, k} + \mat{\Psi}_{k} \bar{\vec{u}}_{i, k} \quad \forall i \in \dist{I} ,\; \forall k \in \dist{K},\label{eq:dynamics_constraint_QCQP}\\
& \bar{\vec{u}}_{i, k} = \vec{0} \quad \forall i \in \dist{I} ,\; \forall k \in \dist{K}_{n}, \label{eq:u0_constraint_QCQP}\\
& \bar{\vec{u}}^{\intercal}_{i, k}\bar{\vec{u}}_{i, k} \leq a_{c}^{2} u^{2}_{i, max} \quad \forall i \in \dist{I} ,\; \forall k \in \dist{K}_{f}, \label{eq:umax_constraint_QCQP}\\
& \begin{multlined}
    \parenth{\bar{\vec{y}}_{i, k} - \bar{\vec{y}}_{j, k}}^{\intercal} \mat{T}_{k}^{\intercal} \mat{T}_{k} \parenth{\vec{y}_{i, k} - \vec{y}_{j, k}} \geq R_\text{CA} \norm{\bar{\vec{y}}_{i, k} - \bar{\vec{y}}_{j, k}} \\
    \hfill \forall i, j \in \dist{I} ,\; i \neq j ,\; \forall k \in \dist{K}, \label{eq:CA_deputy_deputy_QCQP}
\end{multlined}\\
& \bar{\vec{y}}_{i, k}^{\intercal} \mat{T}_{k}^{\intercal} \mat{T}_{k} \vec{y}_{i, k} \geq R_\text{CA} \norm{\bar{\vec{y}}_{i, k}} \quad \forall i \in \dist{I} ,\; \forall k \in \dist{K},\label{eq:CA_deputy_chief_QCQP}
\end{align}
\end{Problem}
where $\bar{\vec{y}}_{i, k}$ is a predicted value for the dimensional ROE vector of the $i^\text{th}$ deputy at time $t_{k}$ and $\norm{\cdot}$ is the second norm of a vector. The solution to Problem \ref{prob:QCQP_formulation} can be obtained sequentially through Sequential Convex Programming (SCP), where $\bar{\vec{y}}_{i, k}$ for the current iteration is set to $\vec{y}_{i, k}$ from the previous iteration. For the first iteration, $\bar{\vec{y}}_{i, k}$ can be obtained in a variety ways, e.g., imposing the dynamics solution, \cref{eq:ROE_dynamics_sol}, from the initial to the final times with no control inputs, or alternatively solving the problem without the collision avoidance constraints first and extracting ${\vec{y}}_{i, k}$ from the solution, then setting $\bar{\vec{y}}_{i, k} = {\vec{y}}_{i, k}$. The SCP scheme is set to terminate when one of the following criteria is satisfied,
\begin{itemize}
    \item $\norm{\bar{\vec{y}}_{i, k} - \vec{y}_{i, k}}\leq\epsilon$ at the current iteration, with $\epsilon>0$ being a user-defined threshold; 
    \item The guidance profile of the current iteration is collision free; 
    \item The user-defined maximum number of iteration is reached, in which case, the solution trajectory is not guaranteed to be collision-free.
\end{itemize}

Since the objective function of Problem \ref{prob:QCQP_formulation}, and that of Problem \ref{prob:nonconvex_formulation} for that matter, aggregates the squared L2 norms of the Delta-V vectors, across deputies and across time instances, it comes as no surprise that the solver gives a stronger emphasis on larger values of Delta-V. While a quadratic objective function suppresses the control peaks, and hence enhances the smoothness of the optimal state trajectory \cite{abdelkarim2023optimization}, it makes Problems \ref{prob:nonconvex_formulation} and \ref{prob:QCQP_formulation} close to being minimization problems for the maximum Delta-V instance. An alternative approach is to set the cost function to the sum of the second-norms of the Delta-V vectors, rather than the sum of the squared norms. Adopting this approach, then transforming the problem into its separable epigraph form \cite{Lobo1998Applications}, which is a very close form to the epigraph problem form \cite{boyd2004convex}, the new problem can be written as a second-order cone orogramming problem as follows:
\begin{Problem}[SOCP formulation]
\label{prob:SOCP_formulation}
\begin{align}
& \min_{\mat{Y}, \bar{\mat{U}}, \mat{\Gamma}} \quad \frac{1}{a_{c}}\sum_{i\in \dist{I}}\sum_{k \in \dist{K}_{f}}{\parenth{\Delta t_{k} \Gamma_{i, k}}} \nonumber\\
& \text{subject to,} \nonumber\\
& \vec{y}_{i, 0} = \vec{y}_{i, 0}, \qquad \vec{y}_{i, m+1} = \vec{y}_{i, f} \quad \forall i \in \dist{I}, \label{eq:boundary_constraints_SOCP}\\
& \vec{y}_{i, k+1} =  \mat{\Phi}_{k} \vec{y}_{i, k} + \mat{\Psi}_{k} \bar{\vec{u}}_{i, k} \quad \forall i \in \dist{I} ,\; \forall k \in \dist{K},\label{eq:dynamics_constraint_SOCP}\\
& \bar{\vec{u}}_{i, k} = \vec{0} \quad \forall i \in \dist{I} ,\; \forall k \in \dist{K}_{n}, \label{eq:u0_constraint_SOCP}\\
& \norm{\bar{\vec{u}}_{i, k}} \leq \Gamma_{i,k}, \quad \forall i \in \dist{I} ,\; \forall k \in \dist{K}_{f}, \label{eq:umax_constraint_SOCP}\\
& \Gamma_{i,k} \leq a_{c} u_{i, max} \quad \forall i \in \dist{I},\; \forall k \in \dist{K}_{f},\\
& \begin{multlined}
    \parenth{\bar{\vec{y}}_{i, k} - \bar{\vec{y}}_{j, k}}^{\intercal} \mat{T}_{k}^{\intercal} \mat{T}_{k} \parenth{\vec{y}_{i, k} - \vec{y}_{j, k}} \geq R_\text{CA} \norm{\bar{\vec{y}}_{i, k} - \bar{\vec{y}}_{j, k}} \\
    \hfill \forall i, j \in \dist{I} ,\; i \neq j ,\; \forall k \in \dist{K}, \label{eq:CA_deputy_deputy_SOCP}
\end{multlined}\\
& \bar{\vec{y}}_{i, k}^{\intercal} \mat{T}_{k}^{\intercal} \mat{T}_{k} \vec{y}_{i, k} \geq R_\text{CA} \norm{\bar{\vec{y}}_{i, k}} \quad \forall i \in \dist{I} ,\; \forall k \in \dist{K},\label{eq:CA_deputy_chief_SOCP}
\end{align}
\end{Problem}
where $\mat{\Gamma}$ is a matrix which collates all the auxiliary variables that had to be introduced in order to put the problem into the separable epigraph form. Formally,
\begin{equation}\label{eq:Gamma_def}
    \mat{\Gamma} = \begin{bmatrix}
        \Gamma_{1, 0} & \Gamma_{2, 0} & \hdots & \Gamma_{N, 0}\\
        \Gamma_{1, 2} & \Gamma_{2, 2} & \hdots & \Gamma_{N, 2}\\
        \vdots & \vdots & \ddots & \vdots\\
        \Gamma_{1, m-1} & \Gamma_{2, m-1} & \hdots & \Gamma_{N, m-1}\\
    \end{bmatrix}.
\end{equation}

It is important to note that SOCP problems can be handled in their native form, i.e., without having to transform the SOC constraints into quadratic ones, only by a handful of solvers, e.g., SCS, ECOS, and MOSEK. Many of the solvers require the SOC constraints to be transformed into quadratic ones, which, if applied to constraint \eqref{eq:umax_constraint_SOCP}, renders it non-convex, unless an additional linear constraint is added, namely, $\Gamma_{i,k} \geq 0 \; \forall i \in \dist{I},\; \forall k \in \dist{K}_{f}$. By adding this additional constraint, many solvers, e.g., Gurobi and Knitro, recognize the transformed quadratic constraint as a second order cone, and treat it as such.\\

Tracking the behaviour of the Second-Order Cone (SOC) constraint, \cref{eq:umax_constraint_SOCP}, is quite interesting since it behaves such that,
\begin{equation}
    \norm{\bar{\vec{u}}^{*}_{i, k}} = \Gamma^{*}_{i,k}, \quad \forall i \in \dist{I} ,\; \forall k \in \dist{K}_{f}, \label{eq:Gamma_equality_SOCP}
\end{equation}
with $\parenth{\cdot}^{*}$ being the optimal solution to the problem, due to the fact that Problem \ref{prob:SOCP_formulation} is pushing $\Gamma_{i,k}$ to be as low as possible (refer to the cost function of the problem), and since the lowest possible value for $\Gamma_{i,k}$ is $\norm{\bar{\vec{u}}_{i, k}}$ according to constraint \eqref{eq:umax_constraint_SOCP}.\\

Problem \ref{prob:SOCP_formulation} can be taken a step further and be formulated as a Linear Programming (LP) problem, which makes it possible for the new formulation to be solved by the vast majority of the numerical optimization solvers.
This transformation can be applied through implementing a piece-wise linearization to the Euclidean norm function in \cref{eq:umax_constraint_SOCP}. Transforming Problem \ref{prob:SOCP_formulation} into an LP one not only enables the possibility to use many solvers that cannot handle the SOCP form, but may also require less time to solve than that of the SOCP formulation, although not necessarily. A less solve time can be anticipated for the LP formulation due to the fact that LP is presumably the simplest form of a convex optimization problem, and also due to the existence of dedicated LP algorithms, e.g., primal and dual simplex, which have matured over the last few decades. However, transforming the SOCP formulation into an LP one, comes at the cost of adding new constraints to the problem.\\

\begin{figure}[ht]
    \centering
    \includegraphics[width=\linewidth]{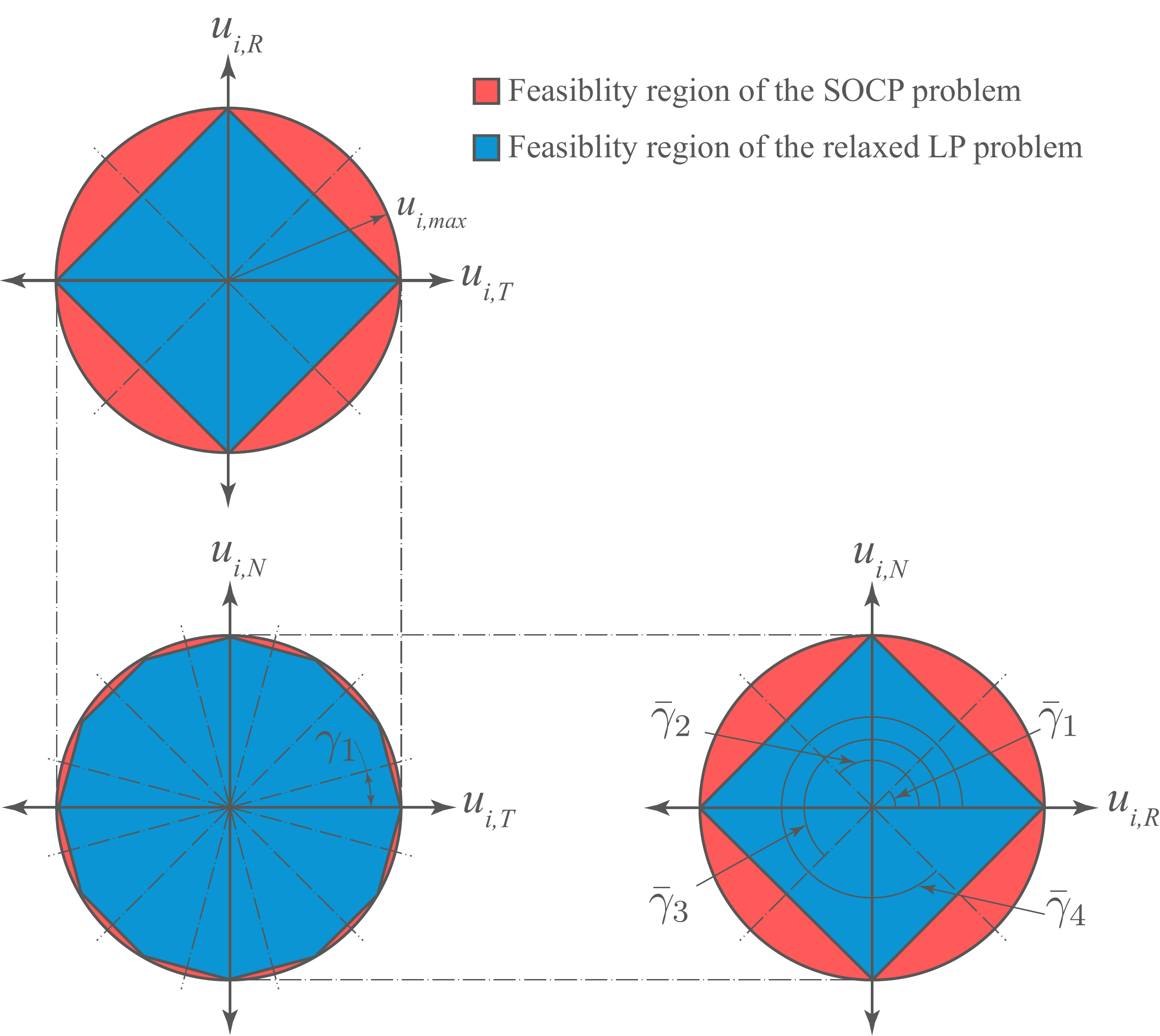}
    \caption{Comparison between the control input feasibility region of the SOCP formulation against that of the LP one}
    \label{fig:feasibility_region_comparison_SOCP_and_LP}
\end{figure}

As previously mentioned, Problem \ref{prob:SOCP_formulation} can be approximated to an LP form through relaxing the maximum acceleration SOCP constraint (sphere), \cref{eq:umax_constraint_SOCP}, using multiple piece-wise affine constraints (polyhedron) as proposed in \cite{Mahfouz2024Fuel-Optimal}. In \cite{Mahfouz2024Fuel-Optimal}, the piece-wise approximation was applied to the sphere such that its projections (circles) on each of the three planes, the Transversal-Normal (T-N), the Radial-Normal(R-N), and the Transversal-Radial(T-R), are each treated separately. Indeed, the optimal solution of Problem \ref{prob:SOCP_formulation} is expected to have minor acceleration components in the radial direction \cite{Gaias2015Impulsive, Belloni2023Relative}, which is why a stronger emphasis is put on the acceleration components lying on the T-N plane by employing a finer grid than that which is applied in the R-N and the T-R planes. This is explained graphically in \cref{fig:feasibility_region_comparison_SOCP_and_LP},  where the feasibility volumes are projected on the T-R, the T-N, and the R-N planes.

Having introduced the concept behind the transformation of Problem \ref{prob:SOCP_formulation} into its LP approximation, this reformulation can be written as follows,
\begin{Problem}[LP formulation]
\label{prob:LP_formulation}
\begin{align}
& \min_{\mat{Y}, \bar{\mat{U}}, \mat{\Gamma}} \quad \frac{1}{a_{c}}\sum_{i\in \dist{I}}\sum_{k \in \dist{K}_{f}}{\parenth{\Delta t_{k} \Gamma_{i, k}}} \nonumber\\
& \text{subject to,} \nonumber\\
& \vec{y}_{i, 0} = \vec{y}_{i, 0}, \qquad \vec{y}_{i, m+1} = \vec{y}_{i, f} \quad \forall i \in \dist{I}, \label{eq:boundary_constraints_LP}\\
& \vec{y}_{i, k+1} =  \mat{\Phi}_{k} \vec{y}_{i, k} + \mat{\Psi}_{k} \bar{\vec{u}}_{i, k} \quad \forall i \in \dist{I} ,\; \forall k \in \dist{K},\label{eq:dynamics_constraint_LP}\\
& \bar{\vec{u}}_{i, k} = \vec{0} \quad \forall i \in \dist{I} ,\; \forall k \in \dist{K}_{n}, \label{eq:u0_constraint_LP}\\
& \begin{multlined}
    \begin{bmatrix} 0 & \cos{\parenth{\gamma_{d}}} & \sin{\parenth{\gamma_{d}}} \end{bmatrix} \bar{\vec{u}}_{i, k} \leq \Gamma_{i,k} \cos{\parenth{\gamma_{max}}}
    \\ \forall d \in \dist{D},\; \forall i \in \dist{I},\; \forall k \in \dist{K}_{f}, \label{eq:umax_constraint_TN_LP}
\end{multlined}\\
& \begin{multlined}
    \begin{bmatrix} \cos{\parenth{\bar{\gamma}_{d}}} & \sin{\parenth{\bar{\gamma}_{d}}} & 0\\
    \cos{\parenth{\bar{\gamma}_{d}}} & 0 & \sin{\parenth{\bar{\gamma}_{d}}}  \end{bmatrix} \bar{\vec{u}}_{i, k} \leq \Gamma_{i, k} \cos{\parenth{\bar{\gamma}_{max}}}\\
    \forall d \in \bar{\dist{D}},\; \forall i \in \dist{I},\; \forall k \in \dist{K}_{f}, \label{eq:umax_constraint_RT_RN_LP}
\end{multlined}\\
& \Gamma_{i,k} \leq a_{c} u_{i, max} \quad \forall i \in \dist{I},\; \forall k \in \dist{K}_{f}, \label{eq:Gamma_LP}\\
& \bar{\vec{y}}_{i, k}^{\intercal} \mat{T}_{k}^{\intercal} \mat{T}_{k} \vec{y}_{i, k} \geq R_\text{CA} \norm{\bar{\vec{y}}_{i, k}} \quad \forall i \in \dist{I},\; \forall k \in \dist{K}.\label{eq:CA_deputy_chief_LP}
\end{align}
\end{Problem}
where $\dist{D} = \left\{1, 2, \hdots, n_\text{dir} \right\}$, with $n_\text{dir} > 4$ being the desired number of affine inequality constraints that approximate the projection of the SOC constraint, \cref{eq:umax_constraint_SOCP}, on the Transversal-Normal plane, $\gamma_{d} = \dfrac{\parenth{2d-1}\pi}{n_\text{dir}}\; \forall d \in \dist{D}$, and $\gamma_{max} = \dfrac{\pi}{n_\text{dir}}$. Moreover, $\bar{\dist{D}} = \left\{1, 2, 3, 4 \right\}$, $\bar{\gamma}_{d} = \dfrac{\parenth{2d-1}\pi}{4} \; \forall d \in \bar{\dist{D}}$, and $\bar{\gamma}_{max} = \dfrac{\pi}{4}$. In \cref{fig:feasibility_region_comparison_SOCP_and_LP}, the constraint relaxations in \cref{eq:umax_constraint_TN_LP} is depicted for $n_\text{dir}=12$ which covers approximately $95.5\%$ of the original constraining circle in the T-N plane.\\

Similar to Problem \ref{prob:SOCP_formulation}, the optimal solution of Problem \ref{prob:LP_formulation} lies at the boundaries of constraints \eqref{eq:umax_constraint_TN_LP} and \eqref{eq:umax_constraint_RT_RN_LP}. This implies that the second norm of the scaled input acceleration, $\norm{\bar{\vec{u}}_{i,k}}$, lies on the surface of the polyhedron (depicted in \cref{fig:feasibility_region_comparison_SOCP_and_LP} for $\Gamma_{i,k} = a_{c} \bar{u}_{i, max}$) which relaxes the original feasibility sphere defined in \cref{eq:umax_constraint_SOCP}.\\

\begin{figure}[ht]
    \centering
    \includegraphics[width=0.7\linewidth]{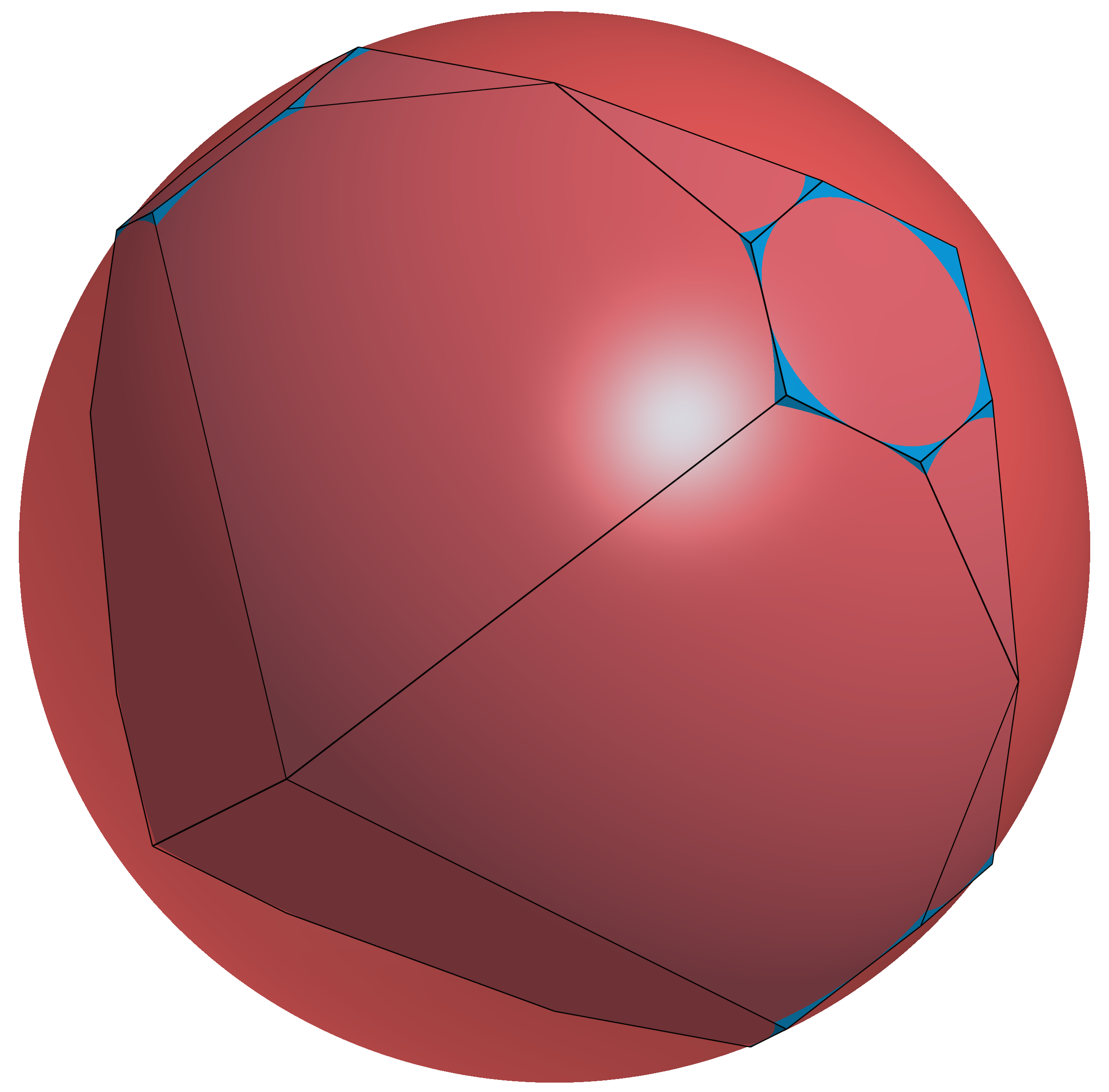}
    \caption{Control input feasibility regions of Problem \ref{prob:SOCP_formulation} and Problem \ref{prob:LP_formulation}}
    \label{fig:Relaxed_feasibility_region_3D}
\end{figure}

It is to be emphasized that while the polyhedron that relaxes the acceleration norm constraint has projections on the T-N, the R-T, and the R-N planes that lie well inside the original feasibility sphere as seen in \cref{fig:feasibility_region_comparison_SOCP_and_LP}, some vertices of the polyhedron can be seen to slightly protrude from this sphere when visualized from different angles in the 3-dimensional space for the case of $n_\text{dir}=12$, which can be seen clearly in \cref{fig:Relaxed_feasibility_region_3D}. Although the relaxing polyhedron of Problem \ref{prob:LP_formulation} can change in size depending on the value of $\Gamma_{i,k}$, and hence lie with its entirety within the original constraining sphere, the fact that it can extend beyond the limits of the sphere when $\Gamma_{i,k}= a_{c} \bar{u}_{i,max}$ can be problematic. As the solution of Problem \ref{prob:LP_formulation} may include points where $\Gamma_{i,k}= a_{c} \bar{u}_{i,max}$, the event of $\norm{\vec{u}_{i,k}}$ getting slightly larger than $u_{i, max}$ becomes a real possibility, since the solution has to lie on the boundaries of constraints \eqref{eq:umax_constraint_TN_LP} and \eqref{eq:umax_constraint_RT_RN_LP} as discussed earlier. 

While the effect of this slight violation of the original constraints can be tolerated for short maneuvers, relative orbit correction maneuvers which require extended periods of time can deviate from their set points by tens or even hundreds of meters in the dimensional ROE space as will be discussed in \cref{sec:Results_and_discussion}. It is for this reason that constraints \eqref{eq:umax_constraint_TN_LP} and \eqref{eq:umax_constraint_RT_RN_LP} are modified such that the approximating polyhedron is uniformly scaled down so that it can be contained entirely by the original sphere regardless of the value of $\Gamma_{i, k}$. The resulting problem after this scaling can be written as,
\begin{Problem}[LP formulation with scaled feas. region]
\label{prob:LP_formulation_scaled}
\begin{align}
& \min_{\mat{Y}, \bar{\mat{U}}, \mat{\Gamma}} \quad \frac{1}{a_{c}}\sum_{i\in \dist{I}}\sum_{k \in \dist{K}_{f}}{\parenth{\Delta t_{k} \Gamma_{i, k}}} \nonumber\\
& \text{subject to,} \nonumber\\
& \vec{y}_{i, 0} = \vec{y}_{i, 0}, \qquad \vec{y}_{i, m+1} = \vec{y}_{i, f} \quad \forall i \in \dist{I}, \label{eq:boundary_constraints_LP_scaled}\\
& \vec{y}_{i, k+1} =  \mat{\Phi}_{k} \vec{y}_{i, k} + \mat{\Psi}_{k} \bar{\vec{u}}_{i, k} \quad \forall i \in \dist{I} ,\; \forall k \in \dist{K},\label{eq:dynamics_constraint_LP_scaled}\\
& \bar{\vec{u}}_{i, k} = \vec{0} \quad \forall i \in \dist{I} ,\; \forall k \in \dist{K}_{n}, \label{eq:u0_constraint_LP_scaled}\\
&\begin{multlined}
    \begin{bmatrix} 0 & \cos{\parenth{\gamma_{d}}} & \sin{\parenth{\gamma_{d}}} \end{bmatrix} c  \bar{\vec{u}}_{i, k} \leq \Gamma_{i,k} \cos{\parenth{\gamma_{max}}}\\
    \forall d \in \dist{D},\; \forall i \in \dist{I},\; \forall k \in \dist{K}_{f}, \label{eq:umax_constraint_TN_LP_scaled}
\end{multlined}\\
& \begin{multlined}
    \begin{bmatrix} \cos{\parenth{\bar{\gamma}_{d}}} & \sin{\parenth{\bar{\gamma}_{d}}} & 0\\
    \cos{\parenth{\bar{\gamma}_{d}}} & 0 & \sin{\parenth{\bar{\gamma}_{d}}}  \end{bmatrix} c \bar{\vec{u}}_{i, k} \leq \Gamma_{i, k} \cos{\parenth{\bar{\gamma}_{max}}}\\
    \forall d \in \bar{\dist{D}},\; \forall i \in \dist{I},\; \forall k \in \dist{K}_{f}, \label{eq:umax_constraint_RT_RN_LP_scaled}
\end{multlined}\\
& \Gamma_{i,k} \leq a_{c} u_{i, max} \quad \forall i \in \dist{I},\; \forall k \in \dist{K}_{f}, \label{eq:Gamma_LP_scaled}\\
& \begin{multlined}
    \parenth{\bar{\vec{y}}_{i, k} - \bar{\vec{y}}_{j, k}}^{\intercal} \mat{T}_{k}^{\intercal} \mat{T}_{k} \parenth{\vec{y}_{i, k} - \vec{y}_{j, k}} \geq R_\text{CA} \norm{\bar{\vec{y}}_{i, k} - \bar{\vec{y}}_{j, k}}\\
    \forall i, j \in \dist{I} ,\; i \neq j ,\; \forall k \in \dist{K}, \label{eq:CA_deputy_deputy_LP_scaled}
\end{multlined}\\
& \bar{\vec{y}}_{i, k}^{\intercal} \mat{T}_{k}^{\intercal} \mat{T}_{k} \vec{y}_{i, k} \geq R_\text{CA} \norm{\bar{\vec{y}}_{i, k}} \quad \forall i \in \dist{I},\; \forall k \in \dist{K},\label{eq:CA_deputy_chief_LP_scaled}
\end{align}
\end{Problem}
where $c$ is a constant scaling factor that guarantees that the polyhedron of \cref{fig:Relaxed_feasibility_region_3D} is situated entirely within the sphere.
The protrusion of the polyhedron beyond the constraining sphere is truly minimal, and could be quantified with the help of a 3D modelling software. In fact, the distance from the center of the polyhedron to the furthest vertex is measured to be approximately $1.7\%$ larger than the radius of the original sphere (for $n_\text{dir}=12$). In the following discussions, the value of $n_\text{dir}$ is set to $12$ while the constant $c$ is fixed to $1.017$.\\

It is important to note that while mission operators usually favor closed-form solutions over numerical optimization-based ones due to the risk of running into an infeasible situation, infeasibility is never a problem in our context. A feasible solution can always be obtained when the maneuver is allowed enough time regardless of how hard the reconfiguration is, as long as the initial and final states are collision-free. Indeed, the maneuver duration is a user-input that can be planned and controlled by the ground operators.

\section{Results and discussion}\label{sec:Results_and_discussion}
In this section, the proposed LP guidance scheme is validated through a case-study and the results of this case-study are analysed. The LP scheme offers more inherent details, such as the scaling factor, for discussion. Moreover, a brief comparison between the convex QCQP, the SOCP, and the relaxed LP formulations is carried out, through a benchmark study of multiple solvers.

\subsection{Case-study}
To test the effectiveness of the guidance schemes, they were run over multiple simulation scenarios. As an example, the details of one of these scenarios are presented here. Considering 4 identical satellites, Sat. A, Sat. B, Sat. C, and Sat. D, that are flying in a close formation with a chief. The four deputies are assumed to be initially equidistant from each other along the transversal direction (trailing/coplanar configuration) with the chief at the center of the formation, and are required to perform a formation reconfiguration such that they are distributed along a Projected Circular Orbit (PCO) at the final time. 
The details of parameterizing the coplanar as well as the PCO configurations in terms of mean ROE is discussed in \cite{scala2021design}, and the initial as well as the final (set-point) states of each of the deputies are summarized in \cref{tab:sim_init_final_states}. The numbers in \cref{tab:sim_init_final_states} indicate that the initial distance between each two consecutive satellites in the trailing formation is $200$ m, while the radius of the final PCO is $300$ m.
Moreover, the orbit of the chief is characterized by $\tilde{\vec{\alpha}}_{c,0} = \begin{bmatrix} 6771 \;\text{km} & 10^{-3} & 98^{\circ} & 0^{\circ} & 0^{\circ} & 180^{\circ} \end{bmatrix}^{\intercal}$ at $t_{0}$.
\begin{table*}[ht]
    \centering
    \caption{Initial and final (required) states for each of the deputies}
    \begin{tabular}{cccc}
    \hline
    \hline
    Satellite & $\vec{y}_{0}\; [\text{m}]$ & $\vec{y}_{f}\; [\text{m}]$ & ~ \\
    \hline
    \rule{0pt}{\tablerowsep}
    Sat. A & $\begin{bmatrix} 0 & -400 & 0 & 0 & 0 & 0 \end{bmatrix}^{\intercal}$ & $\begin{bmatrix} 0 & 0 & 0 & -150 & 300 & 0 \end{bmatrix}^{\intercal}$ & ~ \\
    \rule{0pt}{\tablerowsep}
    Sat. B & $\begin{bmatrix} 0 & -200 & 0 & 0 & 0 & 0 \end{bmatrix}^{\intercal}$ & $\begin{bmatrix} 0 & 0 & -150 & 0 & 0 & -300 \end{bmatrix}^{\intercal}$ & ~ \\
    \rule{0pt}{\tablerowsep}
    Sat. C & $\begin{bmatrix} 0 & 200 & 0 & 0 & 0 & 0 \end{bmatrix}^{\intercal}$ & $\begin{bmatrix} 0 & 0 & 0 & 150 & -300 & 0 \end{bmatrix}^{\intercal}$ & ~ \\
    \rule{0pt}{\tablerowsep}
    Sat. D & $\begin{bmatrix} 0 & 400 & 0 & 0 & 0 & 0 \end{bmatrix}^{\intercal}$ & $\begin{bmatrix} 0 & 0 & 150 & 0 & 0 & 300 \end{bmatrix}^{\intercal}$ & ~ \\
    \hline
    \hline
\end{tabular}
    \label{tab:sim_init_final_states}
\end{table*}

Since the 4 deputies are identical, $u_{i, max}$ are set to $u_{max}$ and $\omega_{i, max}$ are set to $\omega_{max}$ for all $i \in \dist{I}$. Furthermore, the simulation is set up such that there are no operational time constraints, and the durations of the coast arcs are thus set to be all equal, i.e., $T_{n,l} = T_{n} \; \forall l \in \dist{L}$, where $T_{n}$ is calculated according to the equality option in \cref{eq:Tn} for $\omega_{max} = 2\;^{\circ}/\text{s}$ and $T_\text{safety} = 10$ s. The forced motion periods are also fixed to a constant value, i.e., $T_{f,l} = T_{f} \; \forall l \in \dist{L}$, which was tuned around the results of the sensitivity analysis presented in \cite{Mahfouz2024Fuel-Optimal}. A full list of the parameters used in the simulation is given in \cref{tab:guidance_validation_simulation_parameters}. These parameters correspond to that of \href{http://luxspace.lu/resources/}{Triton-X heavy}, designed and manufactured by Luxspace.
{\renewcommand{\arraystretch}{\arraystretchfortable}
\begin{table*}[ht]
    \centering
    \caption{Parameters used in Problem \ref{prob:LP_formulation_scaled} validation simulation}
    \begin{tabular}{cccccccc}
        \hline
        \hline
        {$t_{f}-t_{0}$ \; [\text{orbits}]} & {$T_{f}$\; [\text{orbits}]} & {$T_{n} \; [\text{s}]$} & {$u_{max} \; [\mu\text{m}/\text{s}^{2}]$} & $R_\text{CA} \; [\text{m}]$ & {$n_\text{dir} \; [\text{-}]$} & {$c \; [\text{-}]$}\\
        \hline
        $5$ & $0.2$ & $100$ & $35$ & $100$ & $12$ & $1.017$\\
        \hline
        \hline
    \end{tabular}
\label{tab:guidance_validation_simulation_parameters}
\end{table*}
}

\begin{figure}[ht]
    \centering
    \includegraphics[width=\linewidth]{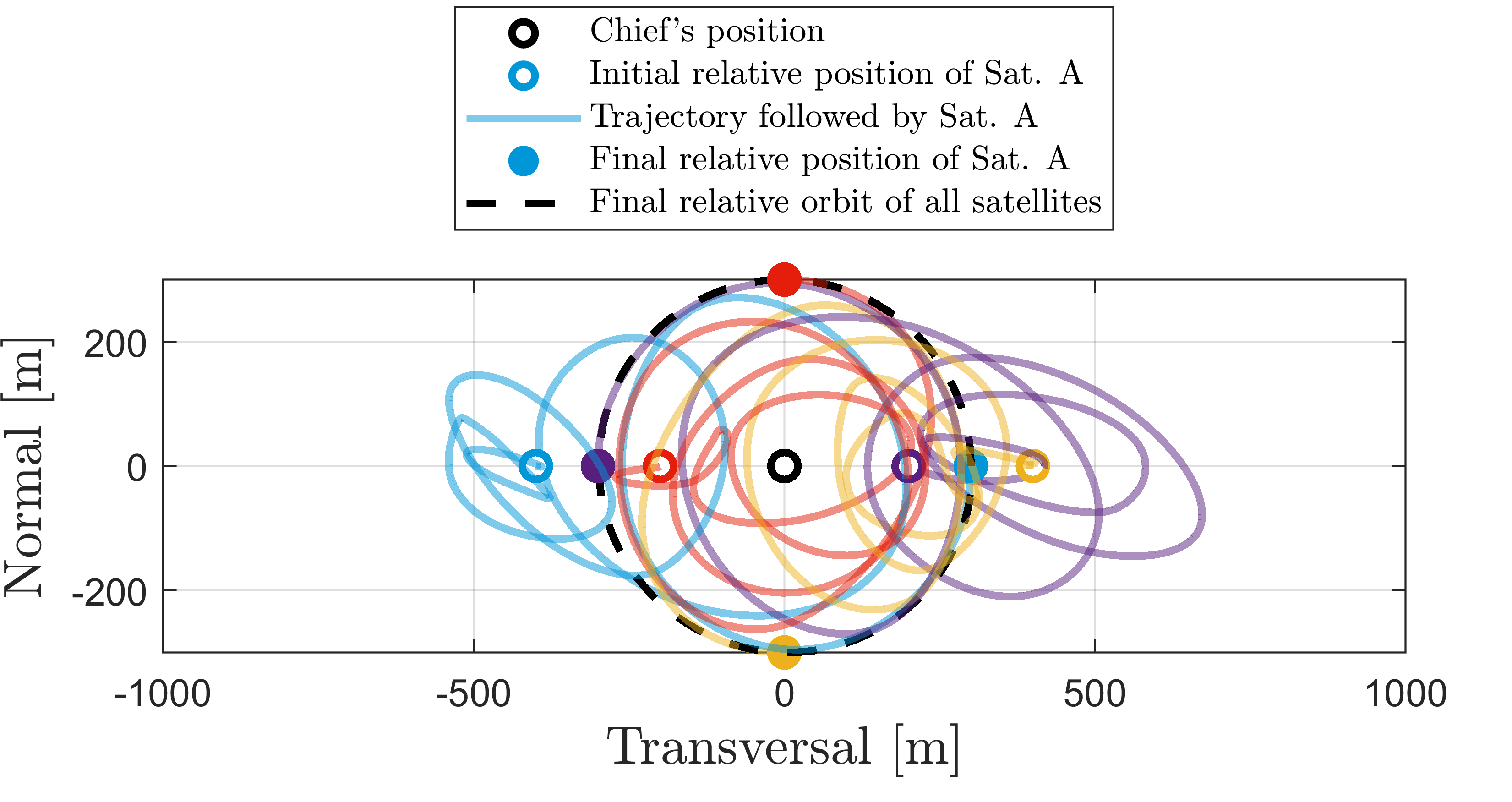}
    \caption{Trajectories followed by each of the deputies throughout the coplanar-to-PCO maneuver}
    \label{fig:Trajectory_of_deputies}
\end{figure}

Running the LP guidance scheme, Problem \ref{prob:LP_formulation_scaled}, over the described reconfiguration setting, profiles for the state vector, the control input vector, and the slack variables are obtained for each deputy at every step of the defined time vector. The trajectory followed by each of the deputies is depicted in \cref{fig:Trajectory_of_deputies}, where the final relative orbit for all the deputies is seen to indeed resemble a PCO with a $300$ m radius. 
Note that \cref{fig:Trajectory_of_deputies} contains legends only for Sat. A to explain the line and marker shapes' convention in use. Legends for other satellites would have had the same shapes as those of Sat. A, yet with their respective colors.

\begin{figure}[ht]
    \centering
    \includegraphics[width=\linewidth]{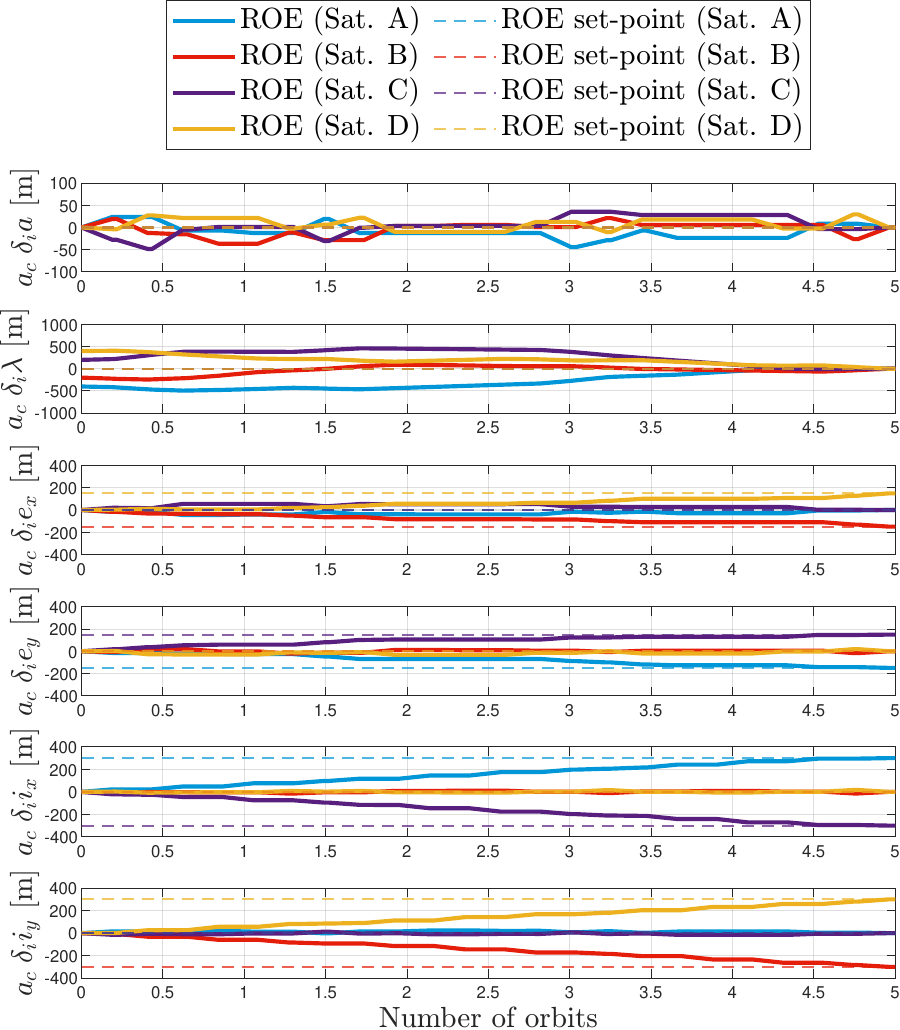}
    \caption{ROE profile over the coplanar-to-PCO maneuver}
    \label{fig:ROE_profile}
\end{figure}

Note that \cref{fig:Trajectory_of_deputies} shows the 2-dimensional view of the trajectory's projection onto the T-N plane. In fact, \cref{tab:sim_init_final_states} suggests that the maneuver is a general one, which requires both, in-plane and out-of-plane corrections. This could be verified by looking either at the 3-dimensional visualization of the trajectories, or at the ROE profile of the maneuver.
The ROE profile of the reconfiguration maneuver is depicted in \cref{fig:ROE_profile}, which not only shows corrections of the in-plane variables as well as the out-of-plane ones, but also demonstrates how each of the relative orbital elements of all the deputies matches its set-point at the final time of the maneuver. 

\begin{figure}[ht]
    \centering
    \includegraphics[width=\linewidth]{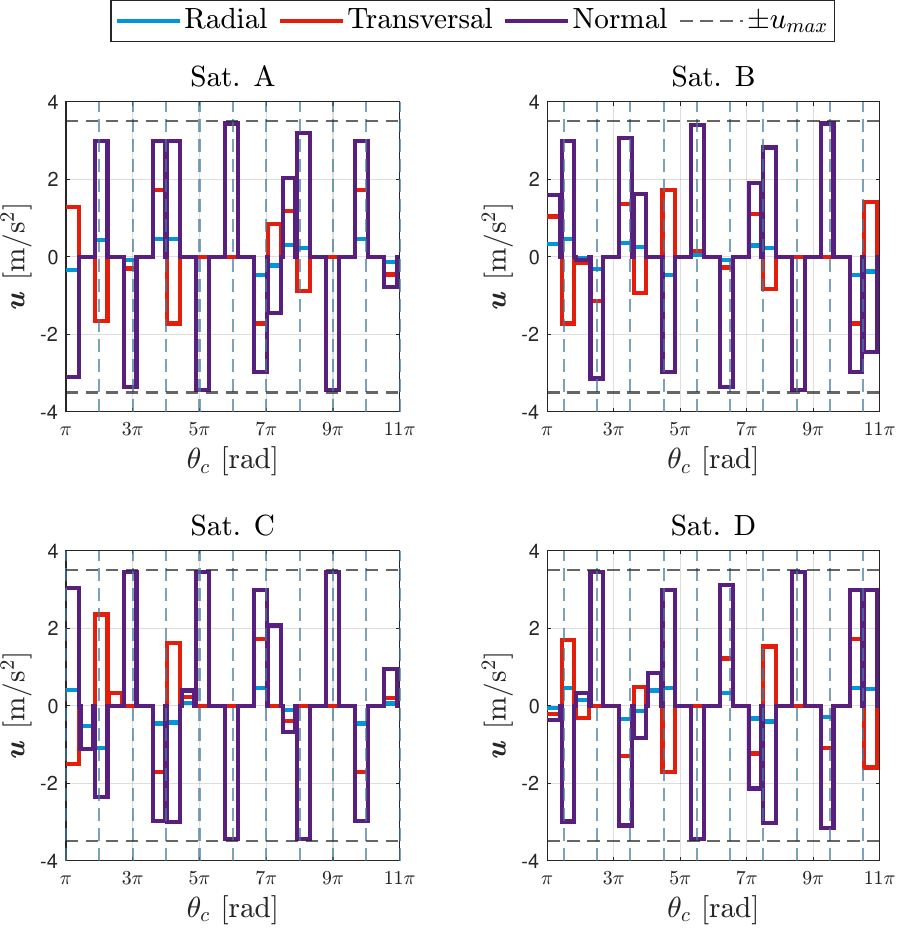}
    \caption{Control acceleration vector profile over the coplanar-to-PCO maneuver}
    \label{fig:Thrust_vector}
\end{figure}

The optimality of the results were investigated through exploring the acceleration profiles of the maneuver. The radial, the transversal, and the normal components of the acceleration provided by each deputy are shown in \cref{fig:Thrust_vector}. 

One interesting aspect which can be seen in \cref{fig:Thrust_vector} is that the radial acceleration component is barely used even though it was not explicitly restricted in the formulation of Problem \ref{prob:LP_formulation_scaled}, simply because optimality dictates not exploiting it. Leveraging mostly the transversal and normal components does indeed coincide with our initial expectations, which formed a foundation for approximating the projections of the maximum acceleration constraint on the R-T and the R-N planes with only 4 affine constraints for each plane, unlike the 12 constraints which approximate the projection of the same constraint on the T-N plane \cite{Mahfouz2024Fuel-Optimal}. Moreover, the fact that the radial acceleration component is minimally utilized suggests that $\delta_{i} \lambda$ corrections are not performed directly through thrust, but rather through varying the value of $\delta_{i} a$ which, in-turn, changes $\delta_{i} \lambda$ according to the natural dynamics \cite{Gaias2015Impulsive}. The small corrections of the relative eccentricity vector are conceivably done using the transversal acceleration component since it is half as expensive as using the radial one from the Delta-V point of view \cite{Gaias2015Impulsive}. Although the in-plane corrections are larger in magnitude, they are mostly done by exploiting the natural dynamics as discussed earlier. Thus, it comes as no surprise that the normal acceleration component, which is responsible only for out-of-plane  corrections, is the most component that has been utilized. In \cref{fig:Thrust_vector}, the dashed teal vertical lines represent the Delta-V-optimal locations to provide impulsive acceleration in the normal direction to achieve the required out-of-plane corrections \cite{Gaias2015Impulsive}. That said, the obtained solution, once again, matches the expectations for an optimal control profile as the normal acceleration component is seen in \cref{fig:Thrust_vector} to act as a bang-bang controller around the optimal locations, while being zero away from them.\\

\begin{figure}[ht]
    \centering
    \captionsetup{justification=centering}
    \includegraphics[width=\linewidth]
    {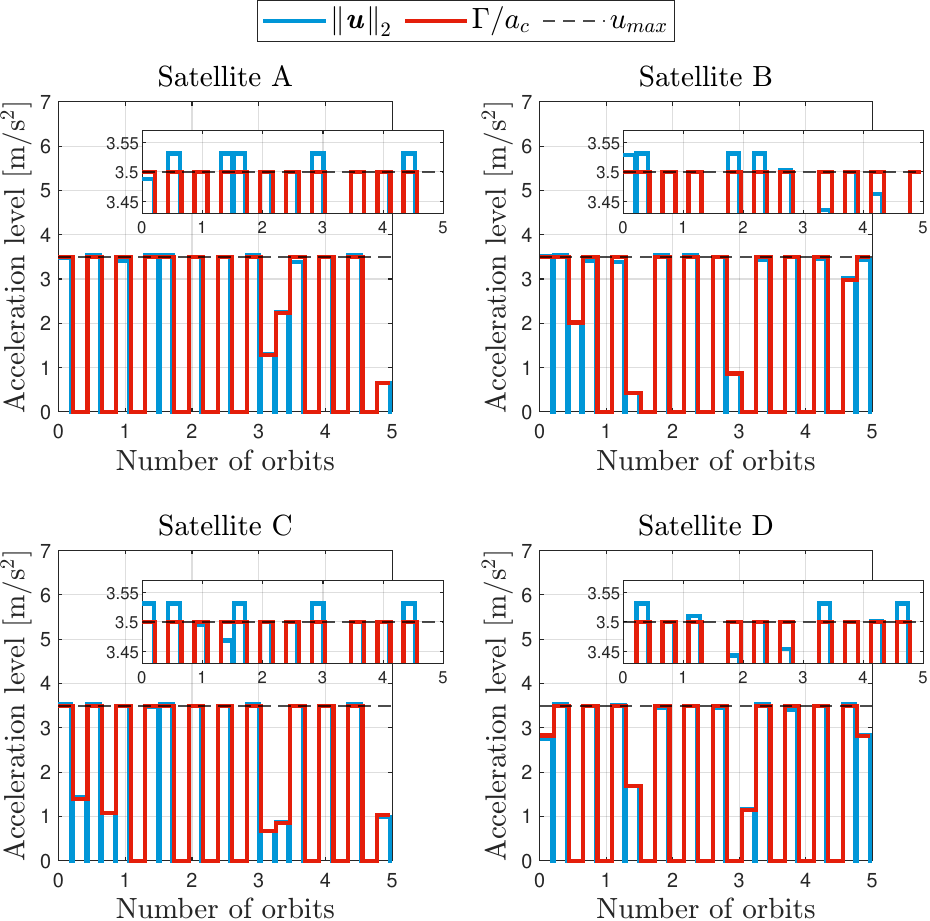}
    \caption{Control acceleration vector and slack variables over the coplanar-to-PCO maneuver, Problem \ref{prob:LP_formulation} (LP formulation)}
    \label{fig:Thrust_norm_gamma_unscaled}
\end{figure}

\begin{figure}[ht]
    \centering
    \captionsetup{justification=centering}
    \includegraphics[width=\linewidth]{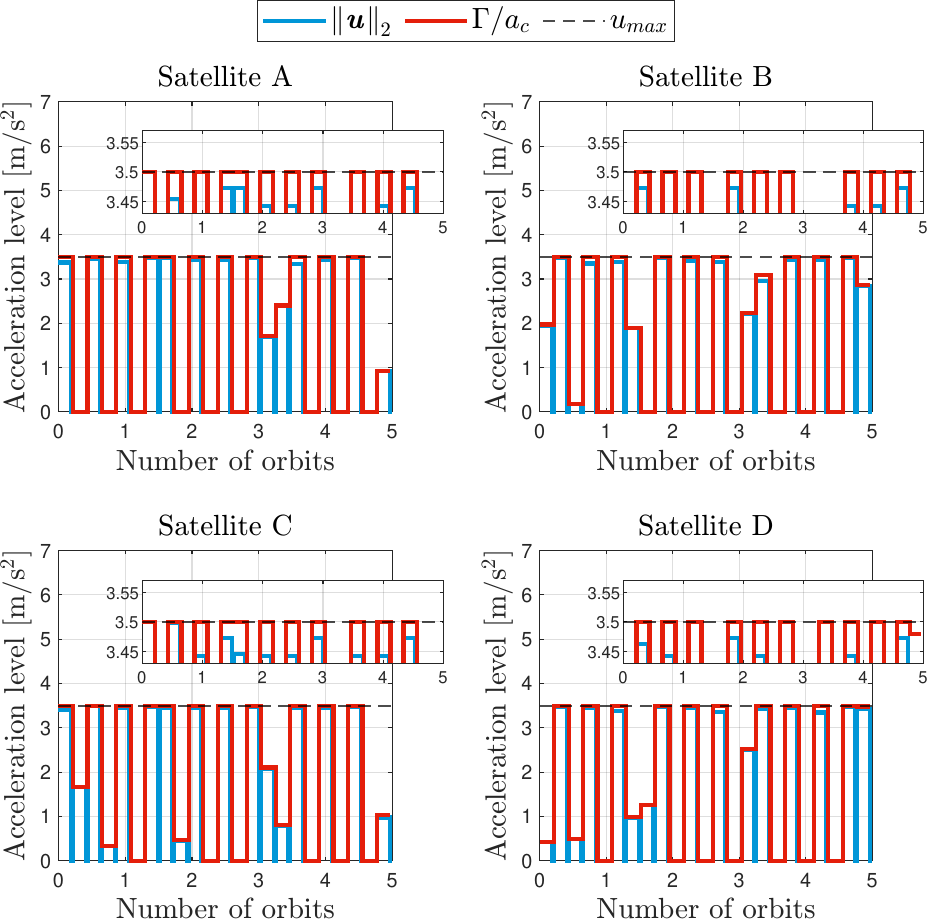}
    \caption{Control acceleration vector and slack variables over the coplanar-to-PCO maneuver, Problem \ref{prob:LP_formulation_scaled} (LP formulation), $c = 1.017$}
    \label{fig:Thrust_norm_gamma_scaled}
\end{figure}

The difference between the two problems, \ref{prob:LP_formulation} and \ref{prob:LP_formulation_scaled}, is only the scaling factor, $c$, which scales down the feasibility region of the maximum acceleration constraint. The effect of this scaling coefficient can be clearly seen by looking into the acceleration (norm) profiles throughout the maneuver, which are depicted in Figures \ref{fig:Thrust_norm_gamma_unscaled} and \ref{fig:Thrust_norm_gamma_scaled} for Problems \ref{prob:LP_formulation} and \ref{prob:LP_formulation_scaled}, respectively.

\begin{figure}[ht]
    \centering
    \captionsetup{justification=centering}
    \includegraphics[width=\linewidth]
    {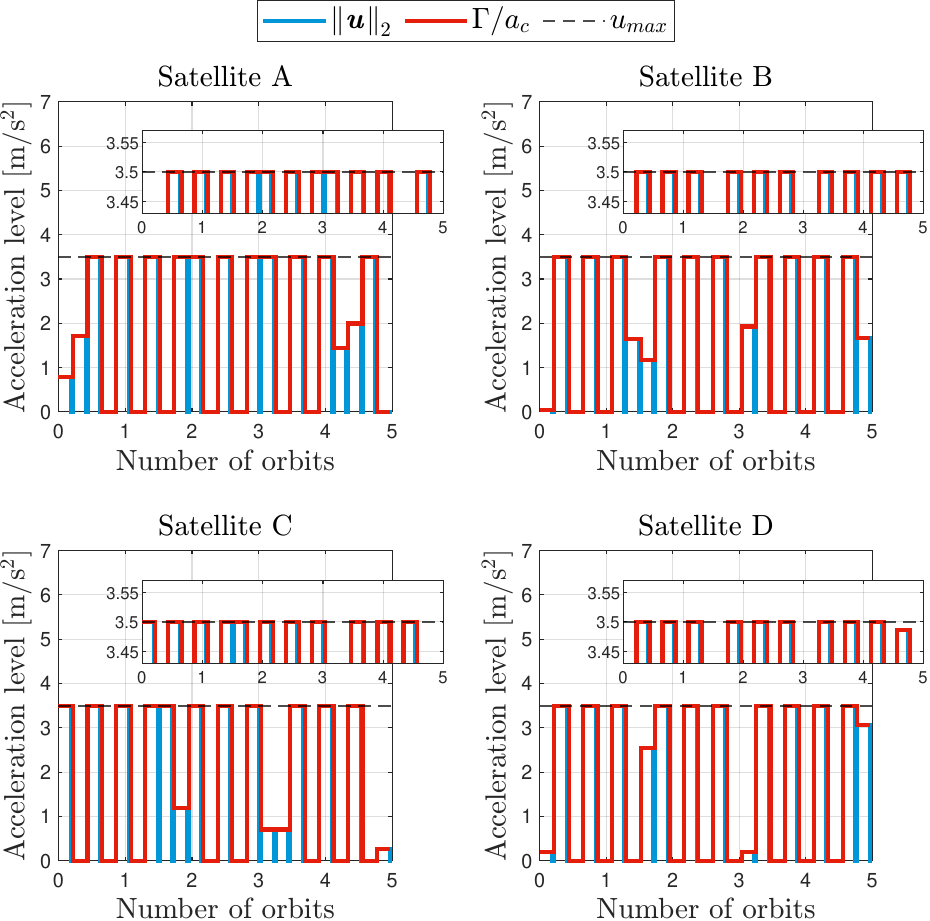}
    \caption{Control acceleration vector and slack variables over the coplanar-to-PCO maneuver, Problem \ref{prob:SOCP_formulation} (SOCP formulation)}
    \label{fig:Thrust_norm_gamma_SOCP}
\end{figure}

When $c$ is set to unity, which is the case in Problem \ref{prob:LP_formulation}, the acceleration is prone surpass the maximum level as seen in \cref{fig:Thrust_norm_gamma_unscaled}. Conversely, 
in \cref{fig:Thrust_norm_gamma_scaled}, our claim that setting $c$ to $1.017$ (for $n_\text{dir} = 12$) guarantees that the L2 norm of the acceleration never exceeds the maximum acceleration level is verified (see the blue line).
It is important to note that in the former case, the acceleration will never surpass the maximum acceleration by more than $1.7\%$ (for $n_\text{dir} = 12$). Despite being minimal, this violation may lead to large errors over time in the ROE space which cannot be simply ignored when thrusters saturation is imposed, especially when the $\delta_{i} a$ variations are used to drive the in-plane variables to their set points.
Figures \ref{fig:Thrust_norm_gamma_scaled} and \ref{fig:Thrust_norm_gamma_unscaled}  also depict a scaled profile of the slack variable, $\Gamma$, throughout the reconfiguration maneuver (red lines), which verifies that constraint \eqref{eq:Gamma_LP_scaled} never gets violated. Note that $\Gamma^{*}_{i,k} \neq a_{c} \norm{\vec{u}^{*}_{i,k}}$ in the two LP formulations of the problem, however $\Gamma^{*}_{i,k} = \norm{\vec{u}^{*}_{i,k}}$ in the SOCP formulation, as stated by \cref{eq:Gamma_equality_SOCP}. This can be verified by inspecting the profiles of the L2 norm of the control acceleration vector when Problem \ref{prob:SOCP_formulation} is solved. These profiles are depicted for the case-study reconfiguration scenario in \cref{fig:Thrust_norm_gamma_SOCP}, which shows that, for all the forced-motion periods, the optimal $\Gamma$ is indeed equal to the acceleration level provided by each deputy.

\begin{figure}[ht]
    \centering
    \captionsetup{justification=centering}
    \includegraphics[width=\linewidth]
    {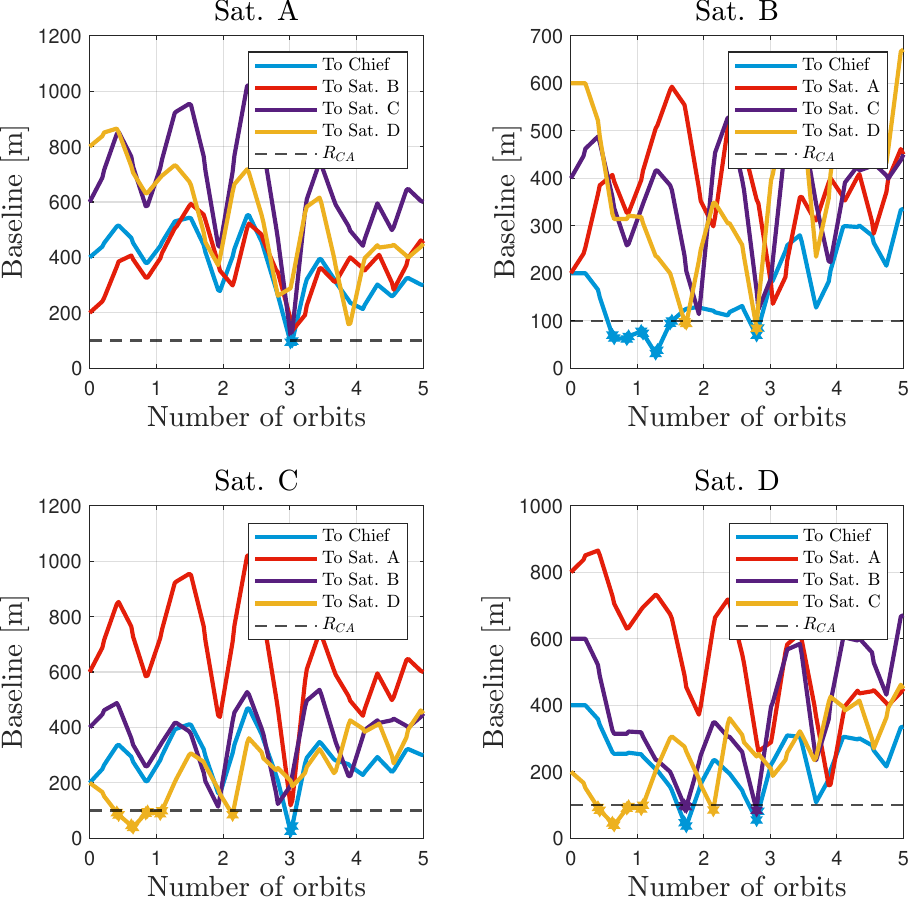}
    \caption{Intersatellite distances over the coplanar-to-PCO maneuver, zeroth iteration}
    \label{fig:intersatellite_distances_zeroth_iteration}
\end{figure}

\begin{figure}[ht]
    \captionsetup{justification=centering}
    \includegraphics[width=\linewidth]{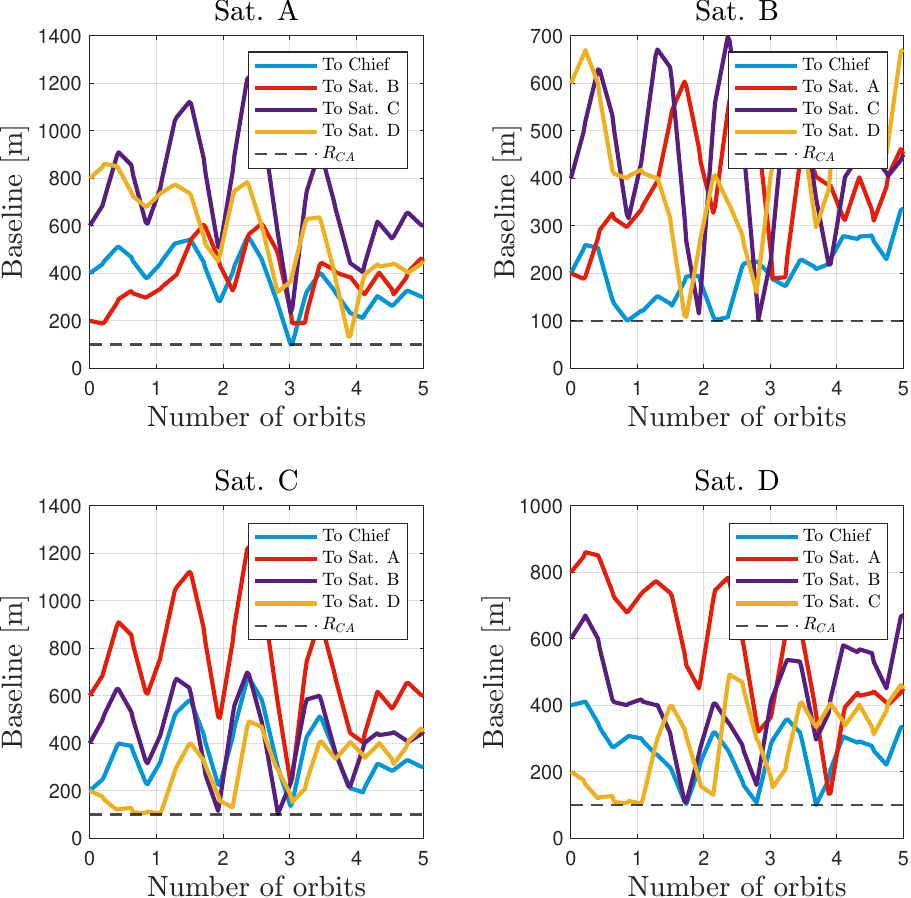}
    \caption{Intersatellite distances over the coplanar-to-PCO maneuver, first iteration}
    \label{fig:intersatellite_distances_first_iteration}
\end{figure}

A very important aspect of the modified guidance schemes, Problems \ref{prob:QCQP_formulation} through \ref{prob:LP_formulation_scaled}, is that they all rely on sequential convex programming, which may require the problem to be solved multiple times before an optimal solution could be obtained. The approach we adopted for solving Problems \ref{prob:QCQP_formulation} through \ref{prob:LP_formulation_scaled} is to solve the problem first without the collision avoidance constraints, inequalities \eqref{eq:CA_deputy_deputy_LP_scaled} and \eqref{eq:CA_deputy_chief_LP_scaled} for Problem \ref{prob:LP_formulation_scaled}, to obtain estimates for $\bar{\vec{y}}_{i,k}$. The problem is subsequently solved iteratively and the values of $\bar{\vec{y}}_{i,k}$ are updated at each iteration until any of the stopping criteria is met. The termination criteria were introduced in \cref{sec:Guidance}, but they are mentioned here once again to allow for a more elaborate discussion. The SCP termination criteria are a) $\norm{\bar{\vec{y}}_{i, k} - \vec{y}_{i, k}}\leq\epsilon$ at the current iteration; b) The guidance profile of the current iteration is collision free; c) The user-defined maximum number of iteration is reached.
Indeed, adopting only the first criterion is guaranteed to result in an optimal guidance profile, if the problem is feasible to begin with, yet at the cost of computational time, since a large number of iterations might be required if $\epsilon$ is chosen to be very small. Implementing the second criterion may lead to a sub-optimal solution, however, it considerably reduces the number of iterations required to solve the problem. The third criterion is nothing but a safeguard to ensure that the solver is not stuck in an infinite loop.
In the context of our case-study, two simulations were run (using the formulation of Problem \ref{prob:LP_formulation_scaled}) where the first adopted the two stopping criteria, a) and c), while the second adopted all of the three. The former required 7 iterations to solve the problem (using $\epsilon=1\;\text{m}$), resulting in a solution which requires a total Delta-V of $1.8\; \text{m}/\text{s}$, while the latter needed only 1 iteration, providing a guidance profile which requires $1.82\; \text{m}/\text{s}$, which calls for only $1\%$ increase in the required Delta-V. One thing which is worth noting is that, throughout our test simulations, adopting the second criterion almost always required a single iteration after the zeroth iteration in which the problem is solved without the collision avoidance constraints. This behaviour is especially expected in cases where the solution of the zeroth iteration is almost collision-free. The intersatellite distances for our case-study are depicted in \cref{fig:intersatellite_distances_zeroth_iteration} for the zeroth SCP iteration and in \cref{fig:intersatellite_distances_first_iteration} for the first SCP iteration, where the asterisks signify a location where collision avoidance is violated.    

\subsection{Solvers benchmark}
The total Delta-V cost of the SOCP formulation, Problem \ref{prob:SOCP_formulation}, is expected to be less than what results from solving the convex QCQP problem, Problem \ref{prob:QCQP_formulation}. It is for this  obvious reason that the QCQP scheme had to be modified into the SOCP formulation. Approximating the SOCP problem by an LP formulation, Problem \ref{prob:LP_formulation_scaled}, could be justified by many motives, among which is the fact that this relaxation puts the problem in the simplest form of convex programming, which is the easiest to implement. Moreover, the LP formulation is expected to require less solve time, especially for lower dimensional reconfiguration scenarios, i.e., those scenarios that involve a low number of sampling instances, and consequently involve a low number of constraints. The LP formulation is not expected to be very fast for higher dimensional reconfiguration scenarios, since one SOCP constraint is approximated by 20 linear constraints; 12 in the T-N plane, 4 in the R-N plane, and 4 in the T-R plane.
One other advantage of transforming an SOCP problem into an approximated linear program is to use dedicated solvers for linear programming that could not be possibly used in an SOCP context. This opens the door to many open-source non-commercial solvers that can handle only linear constraints such as GLPK, HiGHS, OSQP, and many others. Furthermore, some solvers which can handle SOCP problems still use dedicated algorithms for linear and quadratic programs that are not suitable for solving SOCP problems. Examples of such solvers include CPLEX, Gurobi, and Xpress, which usually favor simplex methods for linear programs, while using interior-point algorithms for QCQP and SOCP problems.

To get a clearer insight of the key differences between Problems \ref{prob:QCQP_formulation}, \ref{prob:SOCP_formulation}, and \ref{prob:LP_formulation_scaled}, four different reconfiguration scenarios were identified, and were used for the purpose of benchmarking different solvers when run over each of these formulations. The four scenarios were carefully selected to reflect wide ranges of formation and problem sizes, i.e., intersatellite distances, and number of variables and constraints, while being relevant from the applications point of view. The first reconfiguration requires the formation to go from a pendulum configuration into a PCO, the second reshapes the formation from a PCO into a cartwheel configuration, the third requests a cartwheel to helix reconfiguration, and the fourth starts from a helix configuration and ends up in a pendulum one. The reader is referred to \ref{app:Reconfiguration_scenarios} for the full details of the reconfiguration scenarios. It is to be emphasized that the assumed configuration geometries, i.e., PCO, pendulum, cartwheel, and helix, are all of a great interest for remote sensing applications \cite{Fasano2014Formation_Geometry, Wang2022Optimal_PCO}, and the identified reconfigurations may resemble a formation geometry change within a multi-static Synthetic Aperture Radar (SAR) mission. In the benchmark test, 15 of the most commonly used solvers for the types of problems in hand, according to the Mittelmann's benchmark\footnote{This benchmark used to act a decision tree for optimization software. It can be accessed through: \href{https://plato.asu.edu/bench.html}{https://plato.asu.edu/bench.html}} and according to the statistics of NEOS server\footnote{The statistics of NEOS server can be accessed through: \href{https://neos-server.org/neos/report.html}{https://neos-server.org/neos/report.html}}, were compared. An overview of the adopted 15 solvers is demonstrated in \cref{tab:solvers_overview}, where the problem types that could be handled by a solver are presented, together with information on whether the solver can be freely used in a commercial setting, or a software license needs to be purchased (on the date of writing this study, in June 2024). Note that every QCQP problem can be formulated as an SOCP one \cite{Lobo1998Applications}, nonetheless, the solvers that cannot handle QCQPs in their native form but can handle SOCP problems, e.g., SCS and ECOS, are marked with an \xmark~ mark in the QCQP column. The opposite is not true, however. Solvers which do not accept SOC constraints in their native form but rather need a reformulation of these constraints into the quadratic form are still marked with a \cmark~ mark in the SOCP column, since they still recognize the SOC nature of the reformulated constraints and treat them as such.
\begin{table}[ht]
    \centering
    \caption{Solvers overview}
    \label{tab:solvers_overview}
    \begin{tabular}{lcccc}
    \hline
    \hline
    Solver & Free & QCQP & SOCP & LP\\
    \hline
    \hline
    GLPK \cite{Makhorin2012GLPK} & \cmark & \xmark & \xmark & \cmark \\
    CLP \cite{Forrest2023CLP} & \cmark & \xmark & \xmark & \cmark\\
    OSQP \cite{Stellato2020OSQP} & \cmark & \xmark & \xmark & \cmark\\
    OOQP \cite{Gertz2003OOQP} & \cmark & \xmark & \xmark & \cmark\\
    SCS \cite{ODonoghue2016SCS} & \cmark & \xmark & \cmark & \cmark\\
    ECOS \cite{Domahidi2013ECOS} & \cmark & \xmark & \cmark & \cmark\\
    IPOPT \cite{Wächter2006IPOPT} & \cmark & \cmark & \cmark & \cmark\\
    SCIP \cite{Gleixner2017SCIP} & \cmark & \cmark & \cmark & \cmark\\
    MOSEK \cite{MOSEK} & \xmark\tablefootnote{\label{ft:A.L.}Offers a free academic license} & \cmark & \cmark & \cmark \\
    Gurobi \cite{Gurobi} & \xmark\footref{ft:A.L.} & \cmark & \cmark & \cmark \\
    CPLEX \cite{CPLEX} & \xmark\footref{ft:A.L.} & \cmark & \cmark & \cmark \\
    COPT \cite{Dongdong2023COPT} & \xmark\footref{ft:A.L.} & \cmark & \cmark & \cmark\\
    Knitro \cite{Byrd2006Knitro} & \xmark\tablefootnote{Offers a free academic licenses for the professor and the students during the time of the course} & \cmark & \cmark & \cmark\\
    Xpress \cite{Xpress} & \xmark\tablefootnote{Offers a community license which is limited to a maximum of 5000 variables and constraints} & \cmark & \cmark & \cmark\\
    Matlab \cite{MatlabOptimizationToolbox} & \xmark & \cmark & \cmark  & \cmark\\
    \hline
    \hline
    \end{tabular} 
\end{table}

Using their default parameters, each solver was run 10 times over each reconfiguration scenario using the three problem formulations developed in this paper, i.e., the QCQP formulation, Problem \ref{prob:QCQP_formulation}, the SOCP formulation, Problem \ref{prob:SOCP_formulation}, and the LP formulation, Problem \ref{prob:LP_formulation_scaled}. The average time it took a solver to complete each of the 10 iterations was then recorded. The benchmark was conducted on a Windows-PC which comprises an Intel Core i9-10885H CPU with 16 cores and a clock speed of 2.4 GHz. The language that was used formulate the problems is Matlab, and the problems were passed to each solver either through its native Matlab Application Programming Interface (API). e.g., OSQP and SCS, or through a third-party Matlab interface, e.g., GLPK and CLP. Some interesting details on the 15 adopted solvers and on how they were interfaced with Matlab are presented in \ref{app:solvers_details}. It is of utmost importance to declare that the purpose of this benchmark is not to compare the performance of the selected solvers. After all, tweaking the parameters of a solver may very probably result in a different performance. The purpose is rather to draw some recommendations as to which formulation is better used under which conditions, and also to showcase that using the different proposed formulations for the guidance problem has a major affect on both, the solve time of the problem, and the total Delta-V required for the maneuver. The results of the benchmark are  presented in \cref{tab:Solvers_benchmark}. The table also conveys some of the problem properties which are solver-independent.  Namely the following properties are reported; the number of decision variables, the number of constraints excluding decision variables' bounds, the total Delta-V, and the number of required SCP iterations after the zeroth iterations in order to arrive to a collision-free reconfiguration. The log files for the benchmark experiment are available as supplementary materials to this article.
{\renewcommand{\arraystretch}{\arraystretchfortable}
\begin{table*}[ht]
    \centering
    \caption{Benchmark results}
    \label{tab:Solvers_benchmark}
    \small
    \begin{tabular}{lcccccccccccc}
    \hline
    \hline 
    ~ & \multicolumn{3}{c}{Reconfiguration 1} & \multicolumn{3}{c}{Reconfiguration 2} & \multicolumn{3}{c}{Reconfiguration 3} & \multicolumn{3}{c}{Reconfiguration 4}\\
    \cline{2-13}
    ~ & QCQP & SOCP & LP
    & QCQP & SOCP & LP
    & QCQP & SOCP & LP
    & QCQP & SOCP & LP\\ 
    \hline 
    N. variables & 1248 & 1316 & 1316 & 2412 & 2544 & 2544 & 960 & 1012 & 1012 & 3816 & 4026 & 4026 \\ 
    N. constraints & 1234 & 1234 & 2526 & 2661 & 2661 & 5169 & 676 & 676 & 1664 & 4221 & 4221 & 8211 \\ 
    Total $\Delta V$ [m/s] & 1.18 & 1.03 & 1.05 & 3.04 & 2.66 & 2.76 & 1.32 & 1.25 & 1.31 & 4.77 & 4.30 & 4.50 \\ 
    N. SCP iter. & 1 & 1 & 1 & 1 & 1 & 1 & 0 & 0 & 0 & 1 & 1 & 1 \\  
    Solver & \multicolumn{10}{c}{Solve time [s]} \\
    \hline
    GLPK & - & - & 0.18 & - & - & 0.75 & - & - & 0.03 & - & - & 1.52 \\ 
    CLP & - & - & 0.29 & - & - & 1.77 & - & - & 0.01 & - & - & I.L.\footnotemark[\getrefnumber{ft:I.L.}] \\ 
    OSQP & - & - & 0.50 & - & - & 1.18 & - & - & 0.16 & - & - & 2.12 \\ 
    OOQP & - & - & 0.21 & - & - & 0.46 & - & - & 0.05 & - & - & 0.86 \\ 
    SCS & - & 0.12 & 0.41 & - & 0.14 & 1.07 & - & 0.02 & 0.13 & - & 0.32 & 4.76 \\ 
    ECOS & - & 0.03 & 0.05 & - & 0.08 & 0.12 & - & 0.01 & 0.01 & - & 0.17 & 0.19 \\ 
    IPOPT & 13.39 & 60.24 & 0.85 & 3.80 & 60.26 & 1.93 & 7.19 & 30.11 & 0.26 & 36.13 & 60.32 & 2.88 \\ 
    SCIP & T.L.\footnotemark[\getrefnumber{ft:T.L.}] & 24.47 & 0.26 & T.L.\footnotemark[\getrefnumber{ft:T.L.}] & T.L.\footnotemark[\getrefnumber{ft:T.L.}] & 1.00 & T.L.\footnotemark[\getrefnumber{ft:T.L.}] & 6.20 & 0.09 & T.L.\footnotemark[\getrefnumber{ft:T.L.}] & T.L.\footnotemark[\getrefnumber{ft:T.L.}] & 2.34 \\ 
    MOSEK & 0.06 & 0.05 & 0.07 & 0.14 & 0.11 & 0.18 & 0.01 & 0.01 & 0.01 & 0.17 & 0.18 & 0.28 \\ 
    Gurobi & 0.59 & 0.14 & 0.13 & 0.68 & 0.18 & 0.31 & 0.52 & 0.12 & 0.07 & 1.06 & 0.27 & 0.34 \\ 
    CPLEX & 0.31 & N.I.\footnotemark[\getrefnumber{ft:N.I.}] & 0.30 & 0.40 & N.I.\footnotemark[\getrefnumber{ft:N.I.}] & 0.34 & 0.12 & N.I.\footnotemark[\getrefnumber{ft:N.I.}] & 0.11 & 0.58 & N.I.\footnotemark[\getrefnumber{ft:N.I.}] & 0.46 \\
    COPT & 0.09 & 0.06 & 0.11 & 0.17 & 0.09 & 0.18 & 0.03 & 0.02 & 0.05 & 0.24 & 0.13 & 0.24 \\ 
    Knitro & 0.05 & 0.15 & 0.10 & 0.08 & 0.26 & 0.19 & 0.01 & 0.03 & 0.03 & 0.16 & 0.47 & 0.34 \\ 
    Xpress & 0.17 & 0.16 & 0.15 & L.L.\footnotemark[\getrefnumber{ft:L.L.}] & L.L.\footnotemark[\getrefnumber{ft:L.L.}] & L.L.\footnotemark[\getrefnumber{ft:L.L.}] & 0.07 & 0.07 & 0.04 & L.L.\footnotemark[\getrefnumber{ft:L.L.}] & L.L.\footnotemark[\getrefnumber{ft:L.L.}] & L.L.\footnotemark[\getrefnumber{ft:L.L.}] \\ 
    Matlab & 2.95 & 0.16 & 0.09 & 7.46 & 0.54 & 0.54 & 0.89 & 0.04 & 0.01 & 23.60 & 1.63 & 1.04 \\ 
    \hline 
    \hline 
    \end{tabular}
\end{table*}
}
\stepcounter{footnote}
\footnotetext[\value{footnote}]{\label{ft:I.L.}The solver reached the Iteration Limit (I.L.) before finding a feasible solution}
\stepcounter{footnote}
\footnotetext[\value{footnote}]{\label{ft:T.L.}The solver reached the Time Limit (T.L.) (30 seconds) before finding a feasible solution}
\stepcounter{footnote}
\footnotetext[\value{footnote}]{\label{ft:N.I.}The solver is Not Interfaced (N.I.) for the type of problem in question. Refer to \ref{app:solvers_details} for the reason}
\stepcounter{footnote}
\footnotetext[\value{footnote}]{\label{ft:L.L.}The solver did not solve the problem because of License Limitations (L.L.). The community license of Xpress is limited to a maximum of 5000 variables and constraints}

\cref{tab:Solvers_benchmark} presents, in a quantitative manner, how much of the total Delta-V can be saved simply by adopting the SOCP or the LP formulations. It comes as no surprise that the SOCP formulation is requiring less total Delta-V than the LP formulation, since the SOCP problem is exploring a much larger action space than the LP is (refer to Figures \ref{fig:feasibility_region_comparison_SOCP_and_LP} and \ref{fig:Relaxed_feasibility_region_3D}). It is also clear from \cref{tab:Solvers_benchmark} how the number of variables is identical for the SOCP and the LP formulations, since they both consider all the entries of the $\mat{\Gamma}$ matrix as decision variables. The number of constraints for the LP formulation is always much higher than the identical number of constraints for the QCQP and the SOCP formulations. The reason for this is that each instance of \cref{eq:umax_constraint_SOCP} is approximated by 20 linear constraints, Equations \eqref{eq:umax_constraint_TN_LP_scaled} and \eqref{eq:umax_constraint_RT_RN_LP_scaled}, in the LP formulation. The effect of adopting the various problem formulations on the number of required SCP iterations is unclear, mainly due to the fact that SCP is set to terminate once a collision-free reconfiguration is obtained, which is a sub-optimal approach that turns out to be much faster, as has been pointed out earlier.

\begin{figure}[ht]
    \centering
    \captionsetup{justification=centering}
    \includegraphics[width=\linewidth]
    {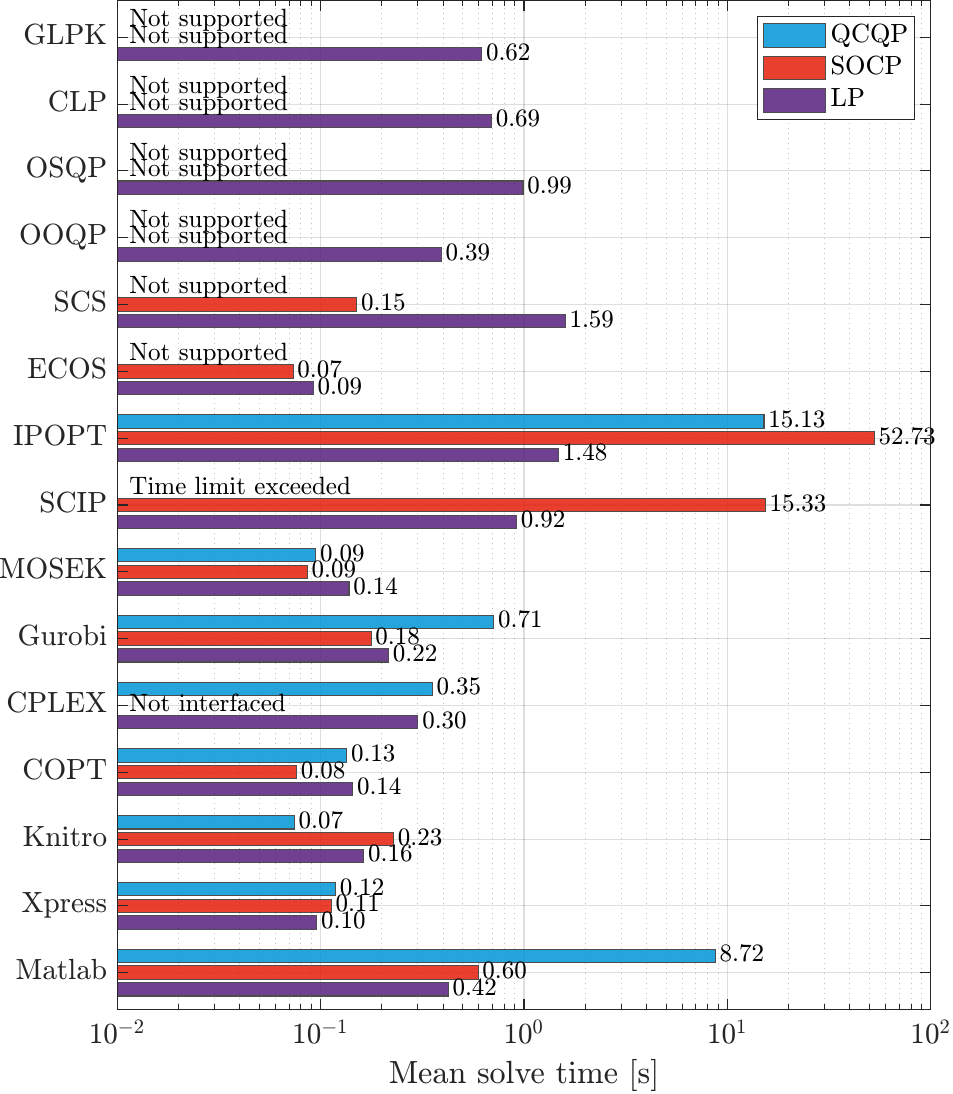}
    \caption{Benchmark summary}
\label{fig:Benchmark_summary}
\end{figure}

To get a better idea of how fast the different solvers return the optimal solution to the problem in hand, the solve time data, reported in \cref{tab:Solvers_benchmark}, were averaged across the four reconfiguration scenarios, and the mean is depicted for each solver in \cref{fig:Benchmark_summary}.

At a first glance on \cref{fig:Benchmark_summary}, it becomes very clear that, among all the solvers, those which are not dedicated to convex optimization, i.e., IPOPT, SCIP, and Matlab in the QCQP case, are the slowest, although IPOPT and SCIP perform competitively in the LP case due to the simplicity of the problem. It is to be noted that many of the benchmarked solvers can still handle Non-Linear Programming (NLP) problems, e.g., CPLEX and Gurobi, however, they do recognize convex optimization problems, and hence, use dedicated algorithms that work best for the convex cases. The LP-only free solvers, i.e., GLPK, CPL, OSQP, and OOQP, present themselves as plausible candidates, although the commercial ones appear to generally perform better for the same problem formulation, with an exception to Matlab's LP solver. Moreover, it is hardly surprising that dedicated conic solvers, namely SCS and ECOS, perform generally faster for the conic case, i.e., SOCP, than for the LP case. In fact, ECOS seems to be performing exceptionally well for a free solver for two problem formulations it supports, i.e., SOCP and LP. 
There is little to no conclusion that can be drawn based on the results of the commercial solvers. For instance, some solvers perform the best for the QCQP formulation, some are favored for the SOCP case, and some are faster when handling the LP problems. The main recommendation from \cref{tab:Solvers_benchmark} as well as \cref{fig:Benchmark_summary} is to avoid the QCQP formulation since it is the most Delta-V intensive, while being generally slower to solve than either of the two other formulations, despite involving the least number of variables and constraints. 
One other takeaway which can be observed by looking into the details of \cref{tab:Solvers_benchmark} is that the performance of the solvers over the LP problems is better when the number of constraints of the original SOCP problem is small, e.g., Reconfiguration 3. This claim is supported by the results of ECOS, MOSEK, and specifically Gurobi. Conversely, for larger problem sizes, the SOCP problems are generally solved faster. It is for this reason that the LP formulation could only be recommended for small problem sizes, while the SOCP formulation is recommended for larger problems, although being harder to implement.\\

\subsection{Limitations of the proposed schemes}
\begin{figure}[ht]
    \centering
    \includegraphics[width=\linewidth]{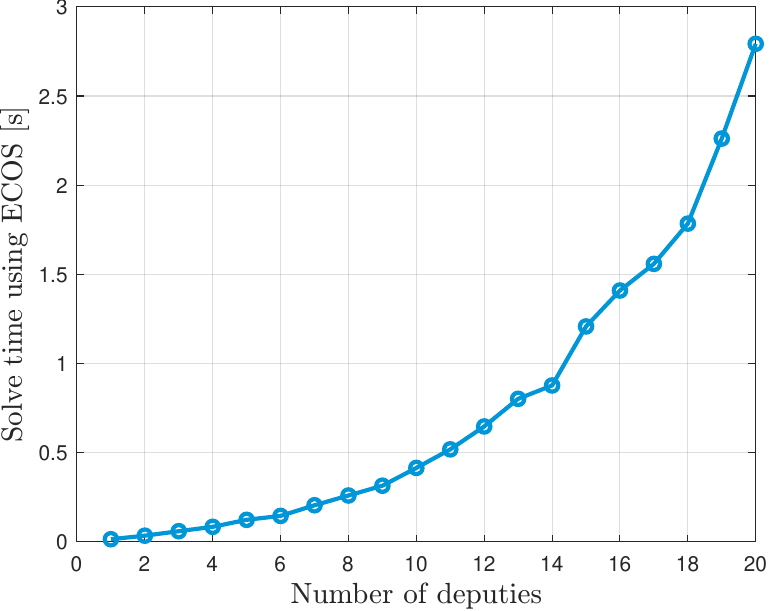}
    \caption{Solve time as the number of deputies increase}
    \label{fig:N_VS_solvetime}
\end{figure}

The proposed guidance strategy can be, theoretically, used for an arbitrary number of deputies. However, employing the proposed guidance schemes in a constellation setting, i.e., with a very large number of satellites, might not be practical from the solve time point of view. To test the limitations of the centralized guidance plan, a last experiment was performed where the number of deputies is allowed to vary from 1 to 20 deputies, while the formation is requested to, once again, perform a Coplanar-to-PCO maneuver. The distance between every two consecutive deputies is set to $200\;\text{m}$ for the initial configuration, while the radius of the final PCO is set $500\;\text{m}$, and, for a fair comparison, each of the $20$ reconfigurations are allowed $10$ orbits to complete.  
Since it has been established that the SOCP formulation is generally recommended, and since ECOS appears to be the fastest solver that handles the SOCP problem in our context, ECOS was run over the $20$ reconfiguration scenarios and the solve time is depicted against the number of deputies in \cref{fig:N_VS_solvetime}.

It is obvious, and also quite conceivable, that the solve time is increasing exponentially with the number of deputies. It is for this reason that the proposed guidance schemes are only recommended for formations with small numbers of deputies. It might be worthwhile, however, to investigate distributed approaches for formations with large numbers of satellites.

\section{Conclusion}\label{sec:Conclusion}
This article proposed centralized guidance schemes for the purpose of reconfiguring the relative orbits of multiple deputies around a chief satellite. One main characteristic of the considered formation reconfiguration is that each deputy is equipped with a single electric thruster, while the chief is uncontrolled, and is treated as the central processing unit for the formation trajectory optimization. In the development of the guidance strategies, the inter-deputy and the deputy-chief collision avoidance is considered, together with the fact that each deputy is under-actuated. The article proposed five different numerical optimization formulations for the guidance problem, where each formulation is a modified/relaxed version of the preceding one. The second (QCQP), the third (SOCP), and the fifth (LP) formulations were identified as the most promising ones, and were involved in an experiment where fifteen solvers were benchmarked accross four different reconfiguration scenarios. It was concluded from the results of the benchmark that the QCQP could never be recommended, as it requires the most total Delta-V change for a maneuver. Moreover, the SOCP formulation is generally recommended, for being fast to solve by many of the benchmarked solvers and also for requiring the least total Delta-V for the maneuver in question. The LP Formulation could only be recommended for small-scale problems, for being very efficient to solve, while being supported by the almost every numerical optimization solver there is.

\section*{Acknowledgments}
This research was funded in whole, or in part, by the Luxembourg National Research Fund (FNR), grant reference BRIDGES/19/MS/14302465. For the purpose of open access, and in fulfilment of the obligations arising from the grant agreement, the author has applied a Creative Commons Attribution 4.0 International (CC BY 4.0) license to any Author Accepted Manuscript version arising from this submission.\\

\appendix
\section{Reconfiguration scenarios used in the benchmark experiment}
\setcounter{figure}{0}
\setcounter{table}{0}
\label{app:Reconfiguration_scenarios}
Four reconfiguration scenarios have been identified for the benchmark experiments. The initial orbit of the chief is assumed to be a sun-synchronous orbit which is shared in all of the four reconfigurations. The chief's orbit is parameterized by $\tilde{\vec{\alpha}}_{c,0} = \begin{bmatrix} 6978 \;\text{km} & 10^{-3} & 97.87^{\circ} & 0^{\circ} & 0^{\circ} & 90^{\circ} \end{bmatrix}^{\intercal}$ at $t_{0}$. Furthermore, the deputies are assumed identical in all the scenarios, and consequently, $u_{i, max}$ are set to $u_{max}$ for all $i \in \curlyb{1, 2, \hdots, N}$ with $N$ being the number of deputies. Furthermore, the durations of the coast arcs are assumed  to be all equal, i.e., $T_{n,l} = T_{n} \; \forall l \in \dist{L}$. The forced motion periods are also fixed to a constant value, i.e., $T_{f,l} = T_{f} \; \forall l \in \dist{L}$. The simulation parameters that are used in the benchmark experiment are the same as those in \cref{tab:guidance_validation_simulation_parameters} and are shared in the four reconfigurations, except for the maneuver duration, which is defined separately for each reconfiguration scenario.

\subsection{Reconfiguration 1 - Pendulum to PCO}
In this reconfiguration scenario, 4 deputy satellites are assumed to be in a pendulum configuration at $t_{0}$ and are required to be reconfigured into a Projected Circular Orbit (PCO) at $t_{f}$, where $t_{f}-t_{0} = 4\; \text{orbits}$. \cref{tab:R1_init_final_states} summarizes the initial and final dimensional ROE vectors, $\vec{y}_{0}$ and  $\vec{y}_{f}$, in meters for all the deputies.
\begin{table*}[ht]
    \centering
    \caption{Initial and final (required) states for each of the deputies in Reconfiguration 1}
    
    \begin{tabular}{ccc}
        \hline
        \hline
        Satellite & $\vec{y}_{0}\; [\text{m}]$ & $\vec{y}_{f}\; [\text{m}]$\\
        \hline
        \rule{0pt}{\tablerowsep}
        
        Sat. A & $\begin{bmatrix} 0& -250& 0& 0& 0& -250 \end{bmatrix}^{\intercal}$ & $\begin{bmatrix} 0& 0& 0& -100& 200& 0 \end{bmatrix}^{\intercal}$ \\
        \rule{0pt}{\tablerowsep}
        
        Sat. B & $\begin{bmatrix} 0& -125& 0& 0& 0& -125 \end{bmatrix}^{\intercal}$ & $\begin{bmatrix} 0& 0& -100& 0& 0& -200 \end{bmatrix}^{\intercal}$\\
        \rule{0pt}{\tablerowsep}
        
        Sat. C & $\begin{bmatrix} 0& 125& 0& 0& 0& 125 \end{bmatrix}^{\intercal}$ &  $\begin{bmatrix} 0& 0& 0& 100& -200& 0 \end{bmatrix}^{\intercal}$\\
        \rule{0pt}{\tablerowsep}
        
        Sat. D & $\begin{bmatrix} 0& 250& 0& 0& 0& 250 \end{bmatrix}$ & 
        $\begin{bmatrix} 0& 0& 100& 0& 0& 200 \end{bmatrix}^{\intercal}$\\
        \hline
        \hline
    \end{tabular}
    \label{tab:R1_init_final_states}
\end{table*}

\begin{figure}[ht]
    \centering
    \includegraphics[width=\linewidth]{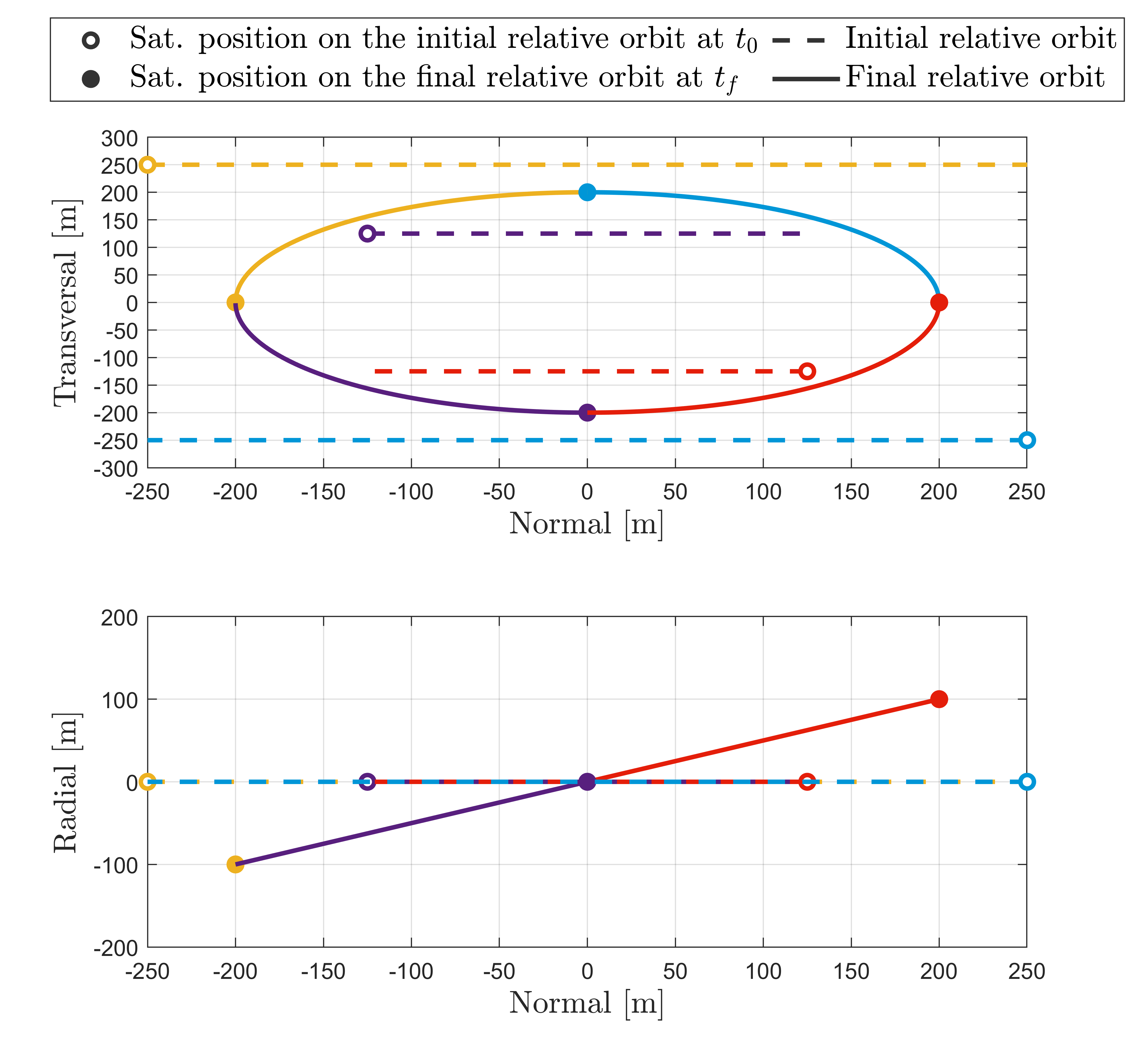}
    \caption{Initial and final orbits of Reconfiguration 1}
    \label{fig:R1_initial_and_final_orbits}
\end{figure}

The shape of the initial and final relative orbits of Reconfiguration 1 is depicted in \cref{fig:R1_initial_and_final_orbits}

\subsection{Reconfiguration 2 - PCO to Cartwheel}
In this reconfiguration scenario, 6 deputy satellites are assumed to be in a PCO configuration at $t_{0}$ and are required to be reconfigured into a cartwheel configuration at $t_{f}$, where $t_{f}-t_{0} = 5\; \text{orbits}$. \cref{tab:R2_init_final_states} summarizes the initial and final dimensional ROE vectors, $\vec{y}_{0}$ and  $\vec{y}_{f}$, in meters for all the deputies.
\begin{table*}[ht]
    \centering
    \caption{Initial and final (required) states for each of the deputies in Reconfiguration 2}
    \begin{tabular}{ccc}
    \hline
    \hline
    Satellite & $\vec{y}_{0}\; [\text{m}]$ & $\vec{y}_{f}\; [\text{m}]$ \\
    \hline
    \rule{0pt}{\tablerowsep}
    
    Sat. A & $\begin{bmatrix} 0 & 0 & 0 & -150 & 300 & 0 \end{bmatrix}^{\intercal}$ & $\begin{bmatrix} 0 & 0 & -500 & 0 & 0 & 0 \end{bmatrix}^{\intercal}$ \\
    \rule{0pt}{\tablerowsep}
    
    Sat. B & $\begin{bmatrix} 0 & -35.91 & -129.90 & -75 & 150 & -259.81 \end{bmatrix}^{\intercal}$ & $\begin{bmatrix} 0 & 0 & -333.33 & 0 & 0 & 0 \end{bmatrix}^{\intercal}$ \\
    \rule{0pt}{\tablerowsep}
    
    Sat. C & $\begin{bmatrix} 0 & -35.91 & -129.90 & 75 & -150 & -259.81 \end{bmatrix}^{\intercal}$ & $\begin{bmatrix} 0 & 0 & -166.67 & 0 & 0 & 0 \end{bmatrix}^{\intercal}$ \\
    \rule{0pt}{\tablerowsep}
    
    Sat. D & $\begin{bmatrix} 0 & 0 & 0 & 150 & -300 & 0 \end{bmatrix}^{\intercal}$ & $\begin{bmatrix} 0 & 0 & 166.67 & 0 & 0 & 0 \end{bmatrix}^{\intercal}$ \\
    \rule{0pt}{\tablerowsep}
    
    Sat. E & $\begin{bmatrix} 0 & 35.91 & 129.90 & 75 & -150 & 259.81 \end{bmatrix}^{\intercal}$ & $\begin{bmatrix} 0 & 0 & 333.33 & 0 & 0 & 0 \end{bmatrix}^{\intercal}$ \\
    \rule{0pt}{\tablerowsep}
    
    Sat. F & $\begin{bmatrix} 0 & 35.91 & 129.90 & -75 & 150 & 259.81 \end{bmatrix}^{\intercal}$ & $\begin{bmatrix} 0 & 0 & 500 & 0 & 0 & 0 \end{bmatrix}^{\intercal}$ \\
    \hline
    \hline
    \end{tabular}
    \label{tab:R2_init_final_states}
\end{table*}

\begin{figure}[ht]
    \centering
    \includegraphics[width=\linewidth]{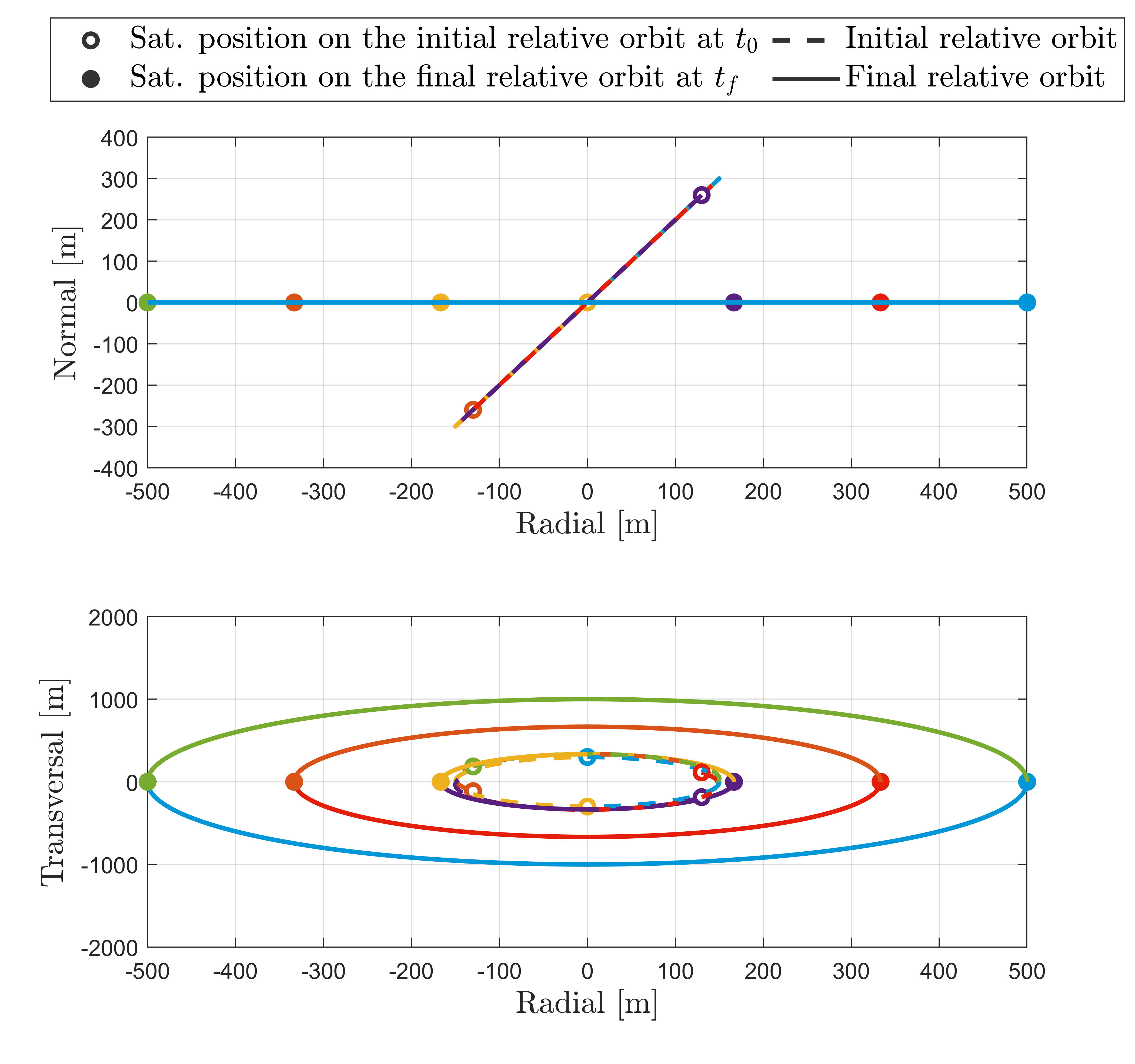}
    \caption{Initial and final orbits of Reconfiguration 2}
    \label{fig:R2_initial_and_final_orbits}
\end{figure}

The shape of the initial and final relative orbits of Reconfiguration 2 is depicted in \cref{fig:R2_initial_and_final_orbits}

\subsection{Reconfiguration 3 - Cartwheel to Helix}
In this reconfiguration scenario, 4 deputy satellites are assumed to be in a cartwheel configuration at $t_{0}$ and are required to be reconfigured into a helix configuration at $t_{f}$, where $t_{f}-t_{0} = 5\; \text{orbits}$. \cref{tab:R3_init_final_states} summarizes the initial and final dimensional ROE vectors, $\vec{y}_{0}$ and  $\vec{y}_{f}$, in meters for all the deputies.
\begin{table*}[ht]
    \centering
    \caption{Initial and final (required) states for each of the deputies in Reconfiguration 3}
    \begin{tabular}{ccc}
    \hline
    \hline
    Satellite & $\vec{y}_{0}\; [\text{m}]$ & $\vec{y}_{f}\; [\text{m}]$ \\
    \hline
    \rule{0pt}{\tablerowsep}
    
    Sat. A & $\begin{bmatrix} 0 & 0 & -500 & 0 & 0 & 0 \end{bmatrix}^{\intercal}$ & $\begin{bmatrix} 0 & 34.56 & -250 & 0 & 0 & -250 \end{bmatrix}^{\intercal}$ \\
    \rule{0pt}{\tablerowsep}
    
    Sat. B & $\begin{bmatrix} 0 & 0 & -250 & 0 & 0 & 0 \end{bmatrix}^{\intercal}$ & $\begin{bmatrix} 0 & 17.28 & -125 & 0 & 0 & -125 \end{bmatrix}^{\intercal}$ \\
    \rule{0pt}{\tablerowsep}
    
    Sat. C & $\begin{bmatrix} 0 & 0 & 250 & 0 & 0 & 0 \end{bmatrix}^{\intercal}$ & $\begin{bmatrix} 0 & -17.28 & 125 & 0 & 0 & 125 \end{bmatrix}^{\intercal}$ \\
    \rule{0pt}{\tablerowsep}
    
    Sat. D & $\begin{bmatrix} 0 & 0 & 500 & 0 & 0 & 0 \end{bmatrix}^{\intercal}$ & $\begin{bmatrix} 0 & -34.56 & 250 & 0 & 0 & 250 \end{bmatrix}^{\intercal}$ \\
    \hline
    \hline
    \end{tabular}
    \label{tab:R3_init_final_states}
\end{table*}

\begin{figure}[ht]
    \centering
    \includegraphics[width=\linewidth]{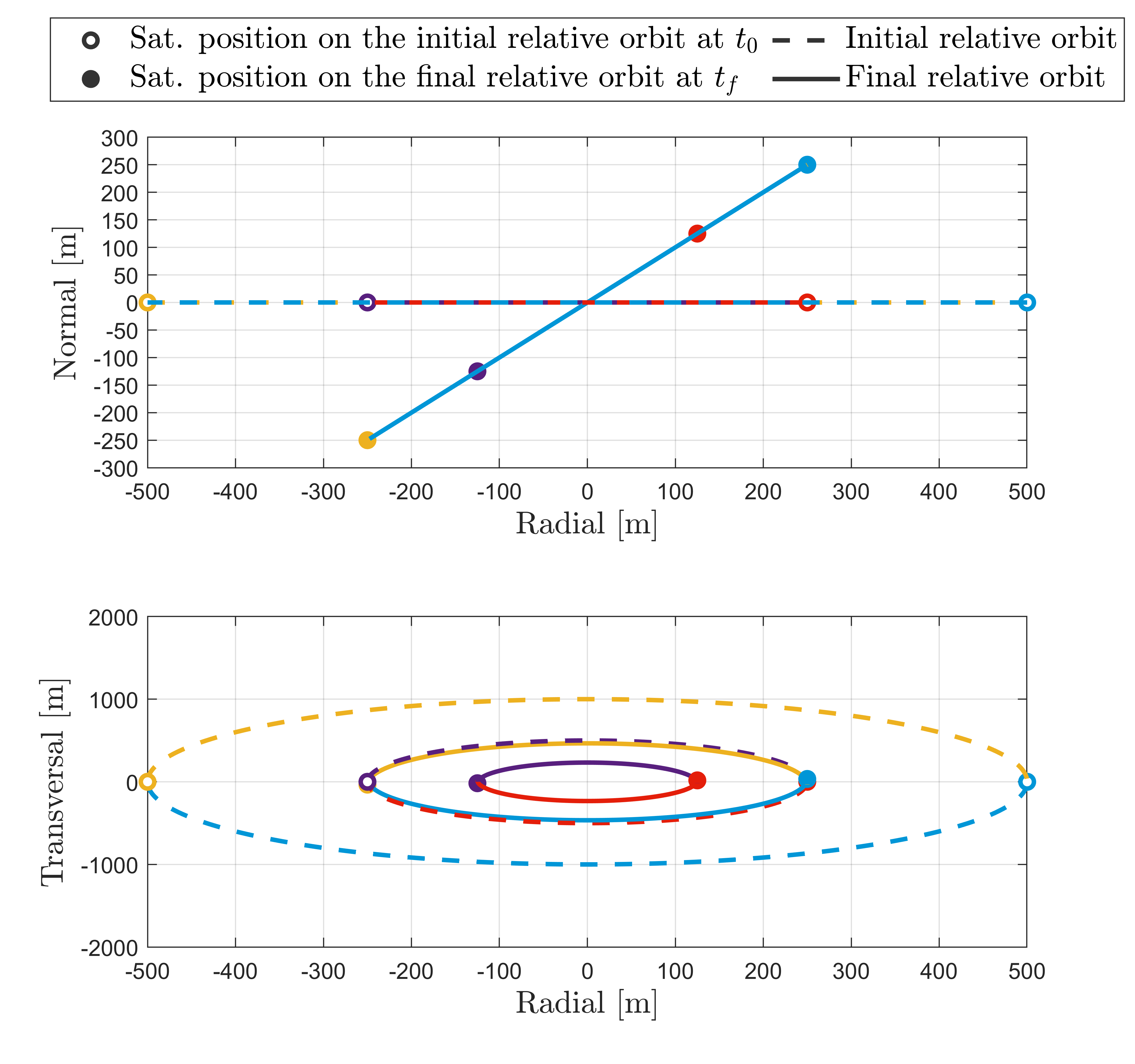}
    \caption{Initial and final orbits of Reconfiguration 3}
    \label{fig:R3_initial_and_final_orbits}
\end{figure}

The shape of the initial and final relative orbits of Reconfiguration 3 is depicted in \cref{fig:R3_initial_and_final_orbits}

\subsection{Reconfiguration 4 - Helix to Pendulum}
In this reconfiguration scenario, 6 deputy satellites are assumed to be in a helix configuration at $t_{0}$ and are required to be reconfigured into a pendulum configuration at $t_{f}$, where $t_{f}-t_{0} = 8\; \text{orbits}$. \cref{tab:R4_init_final_states} summarizes the initial and final dimensional ROE vectors, $\vec{y}_{0}$ and  $\vec{y}_{f}$, in meters for all the deputies.
\begin{table*}[ht]
    \centering
    \caption{Initial and final (required) states for each of the deputies in Reconfiguration 4}
    \begin{tabular}{ccc}
    \hline
    \hline
    Satellite & $\vec{y}_{0}\; [\text{m}]$ & $\vec{y}_{f}\; [\text{m}]$ \\
    \hline
    \rule{0pt}{\tablerowsep}
    
    Sat. A & $\begin{bmatrix} 0 & 34.56 & -250 & 0 & 0 & -250 \end{bmatrix}^{\intercal}$ & $\begin{bmatrix} 0 & -1000 & 0 & 0 & 0 & -1000 \end{bmatrix}^{\intercal}$ \\
    \rule{0pt}{\tablerowsep}
    
    Sat. B & $\begin{bmatrix} 0 & 23.04 & -166.67 & 0 & 0 & -166.67 \end{bmatrix}^{\intercal}$ & $\begin{bmatrix} 0 & -666.67 & 0 & 0 & 0 & -666.67 \end{bmatrix}^{\intercal}$ \\
    \rule{0pt}{\tablerowsep}
    
    Sat. C & $\begin{bmatrix} 0 & 11.52 & -83.33 & 0 & 0 & -83.33 \end{bmatrix}^{\intercal}$ & $\begin{bmatrix} 0 & -333.33 & 0 & 0 & 0 & -333.33 \end{bmatrix}^{\intercal}$ \\
    \rule{0pt}{\tablerowsep}
    
    Sat. D & $\begin{bmatrix} 0 & -11.52 & 83.33 & 0 & 0 & 83.33 \end{bmatrix}^{\intercal}$ & $\begin{bmatrix} 0 & 333.33 & 0 & 0 & 0 & 333.33 \end{bmatrix}^{\intercal}$ \\
    \rule{0pt}{\tablerowsep}
    
    Sat. E & $\begin{bmatrix} 0 & -23.04 & 166.67 & 0 & 0 & 166.67 \end{bmatrix}^{\intercal}$ & $\begin{bmatrix} 0 & 666.67 & 0 & 0 & 0 & 666.67 \end{bmatrix}^{\intercal}$ \\
    \rule{0pt}{\tablerowsep}
    
    Sat. F & $\begin{bmatrix} 0 & -34.56 & 250.00 & 0 & 0 & 250.00 \end{bmatrix}^{\intercal}$ & $\begin{bmatrix} 0 & 1000.00 & 0 & 0 & 0 & 1000.00 \end{bmatrix}^{\intercal}$ \\
    \hline
    \hline
    \end{tabular}
    \label{tab:R4_init_final_states}
\end{table*}

\begin{figure}[ht]
    \centering
    \includegraphics[width=\linewidth]{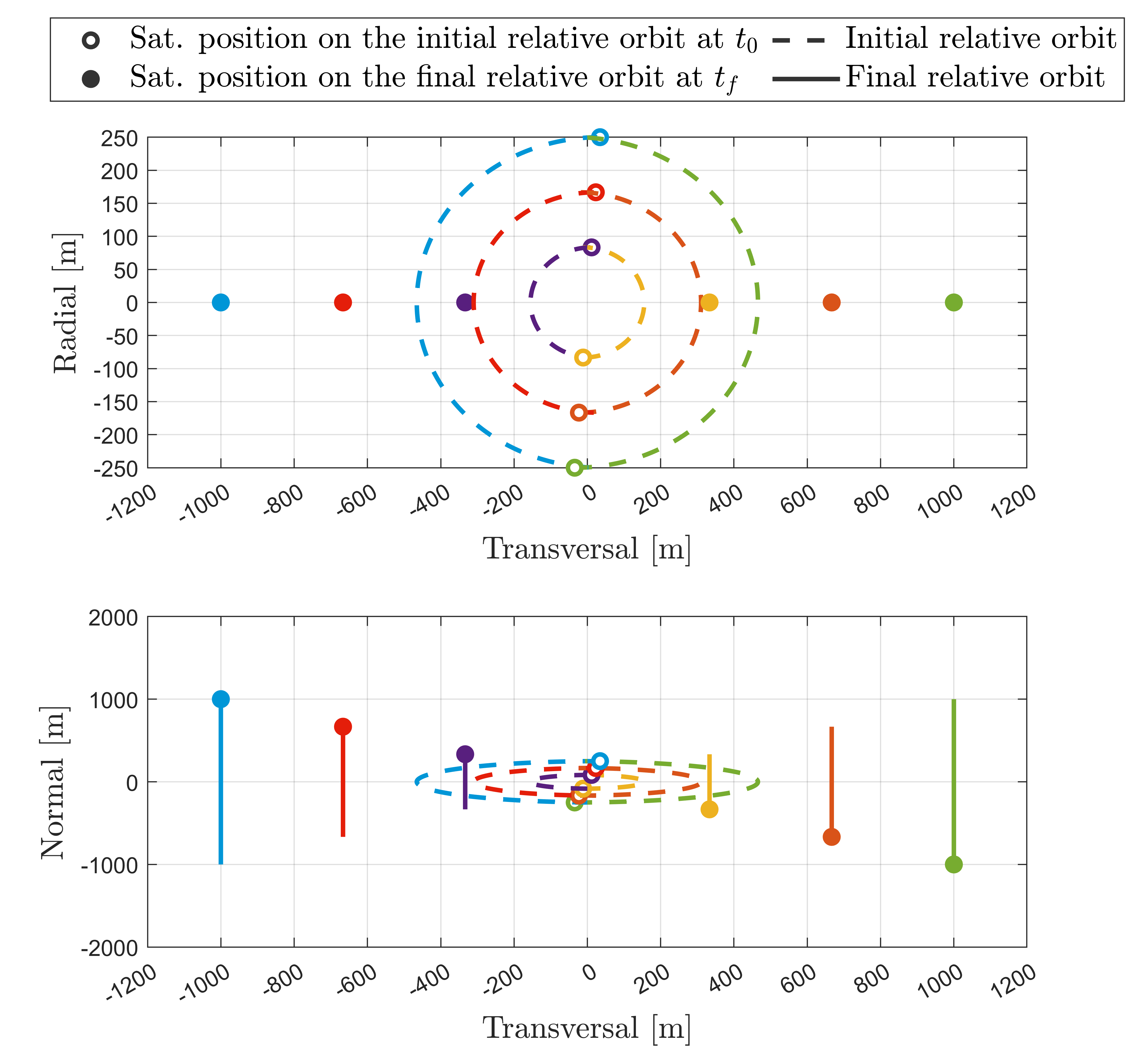}
    \caption{Initial and final orbits of Reconfiguration 4}
    \label{fig:R4_initial_and_final_orbits}
\end{figure}

The shape of the initial and final relative orbits of Reconfiguration 4 is depicted in \cref{fig:R4_initial_and_final_orbits}

\section{Solvers details}
\setcounter{table}{0}
\setcounter{figure}{0}
\label{app:solvers_details}
Fifteen solvers were chosen for the benchmark using the four reconfiguration scenarios defined in \ref{app:Reconfiguration_scenarios}. These solvers were chosen from the list of the most commonly used solvers on NEOS server in the category of LP and SOCP problems, and also from the solvers involved in the Mittelmann's benchmark. Since NEOS server does not have a distinct category for QCQP, the solvers which are used for the SOCP problems are considered, since SOCP is the closest form of a convex programming formulation to the QCQP, and since any QCQP problem can be eventually formulated as an SOCP one. On top of the common solvers for each category, IPOPT was added for being one of the most commonly used free solvers for a variety of problem classes. Moreover, the solvers of the Matlab Optimization Toolbox were added to the list of solvers that are involved in our benchmark experiment under the name "Matlab". A brief overview of the 15 solvers is presented in the following bullet points:

\begin{itemize}
    \item {GLPK (GNU Linear Programming Kit)}: GLPK is intended for solving large-scale linear programming, mixed integer programming (MIP), and other related problems.
    \item {CLP (COIN-OR Linear Program solver)}: CLP is part of the COIN-OR project and is a solver meant for linear programming problems.
    \item {OSQP (Operator Splitting Quadratic Program solver)}: OSQP is a numerical optimization package for solving convex quadratic programs. Linear programs, as a subset of quadratic programming problems, can be handled by the solver.
    \item {OOQP (Object-Oriented software for Quadratic Programs)}: OOQP is an object-oriented software package for solving convex quadratic programming problems.
    \item {SCS (Splitting Conic Solver)}: SCS is primarily meant for large-scale convex quadratic cone problems.
    \item {ECOS (Embedded Conic Solver)}: ECOS is a numerical software for solving convex second-order cone programs.
    \item {IPOPT (Interior Point OPTimizer)}: IPOPT is vastly used for large scale nonlinear optimization of continuous systems.
    \item {SCIP (Solving Constraint Integer Programs)}: SCIP is one of the fastest non-commercial solvers for mixed integer programming and mixed integer nonlinear programming (MINLP). SCIP uses the SoPlex solver internally to solve linear programming problems.
    \item {MOSEK}: MOSEK is a commercial solver with an emphasis on solving large-scale sparse problems. It supports a wide range of problem types, including linear, quadratic, and convex nonlinear programs (NLP).
    \item {Gurobi}: Gurobi is a commercial solver that performs very well in large-scale optimization settings. It can solve various types of optimization problems, including linear as well as nonlinear programming problems. 
    \item {IBM ILOG CPLEX Optimization Studio (CPLEX)}: CPLEX is high-performance optimization solver for linear, mixed-integer and quadratic programming. It was named the for the simplex method implemented in the C programming language, although, at the moment, it supports other types of mathematical optimization and offers interfaces other than C.
    \item {COPT (Cardinal optimizer)}: COPT is a mathematical optimization solver for large-scale problems. It supports many problem types including LP, SOCP, and Convex QCQP.
    \item {Knitro}: Knitro is a commercial software package for solving large scale nonlinear mathematical optimization problems. It supports a wide range of problem types, including linear and nonlinear programs, including non-convex NLPs.
    \item {Xpress}: Xpress is a mathematical optimization solver designed to solve linear programming, mixed integer programming, and other types of optimization problems.
    \item {Matlab}: Matlab is a high-level language and interactive environment that is used to perform computationally intensive tasks. In our context "Matlab" refers to the solvers offered by the Matlab Optimization Toolbox, which includes a dedicated linear programming solver, \code{linprog}, a dedicated conic programming function, \code{coneprog}, a general purpose NLP solver, \code{fmincon}, and many other solvers and capabilities.
\end{itemize}

Matlab has been used as a modelling language, and the problems were passed to each solver through an interface, either provided by the solver developer, by Matlab, or by a third party. \cref{tab:Benchmark_solvers_information} contains the information about the version of each solver used in the benchmark, the interface that has been used between Matlab and the solver.
{\renewcommand{\arraystretch}{\arraystretchfortable}
\begin{table}[ht]
    \centering
    \caption{Benchmark solvers information}
    \begin{tabular}{lcc}
        \hline
        \hline
        Solver & Version & Interface with Matlab\\
        \hline
        GLPK & v4.62 & GLPKmex\tablefootnote{The details of the interface as well as its source files are available on:\href{https://github.com/blegat/glpkmex}{https://github.com/blegat/glpkmex}} \\
        CLP & v1.16.11 & OPTI\tablefootnote{\label{ft:OPTI}The details of the interface as well as its source files are available on:\href{https://github.com/jonathancurrie/OPTI}{https://github.com/jonathancurrie/OPTI}} \\
        OSQP & v0.6.2 & Developer's API \\
        OOQP & v0.99.22 & OPTI\footref{ft:OPTI} \\
        SCS & v3.2.3 & SCS-Matlab\tablefootnote{The details of the interface as well as its source files are available on:\href{https://github.com/bodono/scs-matlab}{https://github.com/bodono/scs-matlab}} \\
        ECOS & v2.0.10 & Developer's API \\
        IPOPT & v3.14.4 & mexIPOPT\tablefootnote{The details of the interface as well as its source files are available on:\href{https://github.com/ebertolazzi/mexIPOPT}{https://github.com/ebertolazzi/mexIPOPT}} \\
        SCIP & v5.0.1 & OPTI\footref{ft:OPTI}\\
        MOSEK & v10.1.15 & Developer's API\\
        Gurobi & v11.0.2 & Developer's API\\
        CPLEX & v12.10.0.0 & OPTI\footref{ft:OPTI} \\
        COPT & v7.1.3 & COPT-MATLAB\tablefootnote{The details of the interface as well as its source files are available on:\href{https://github.com/leavesgrp/COPT-MATLAB}{https://github.com/leavesgrp/COPT-MATLAB}}\\
        Knitro & v14.0.0 & Developer's API\\
        Xpress & v9.4.0 & Developer's API\\
        Matlab & v9.2 (R2021b) & No interface needed\\
        \hline
        \hline
    \end{tabular}
    \label{tab:Benchmark_solvers_information}
\end{table}
}
The most interesting interfacing case is that of CPLEX, which is interfaced through the OPTI Matlab toolbox \cite{Currie2012OPTI}. Although International Business Machines Corporation (IBM) used to provide a Matlab API for their CPLEX solver, this interface was discontinued, and the latest compatible Matlab version with that API was R2019b. Since the OPTI Matlab toolbox provides an interface that works flawlessly with newer Matlab versions, and since the authors used a newer Matlab version to model the problems, with no guarantee that the same code can be used in Matlab2019b to produce the parameters of each problem, it was decided to interface CPLEX with Matlab through OPTI, instead of moving to an older version of Matlab. It is to be noted, however, that SOCP problems are not handled by OPTI.
Another thing to note is that most solvers have multiple internal methods that the user can choose from according to the problem in hand. Many solvers offer the possibility to choose the most suitable algorithm for the problem in hand, according to the pre-solve results. In the benchmark experiment, the algorithm was set to the default, which in most cases means that the solve method is automatically chosen by the solver. The specific algorithms used in the benchmark experiment are reported, by most solvers, in the benchmark log files, which are available as supplementary materials to this article.

 \bibliographystyle{elsarticle-num} 
 \bibliography{References.bib}
\end{document}